\documentclass[12pt]{article}

\def\cstok#1{\leavevmode\thinspace\hbox{\vrule\vtop{\vbox{\hrule\kern1pt
\hbox{\vphantom{\tt/}\thinspace{\tt#1}\thinspace}}
\kern1pt\hrule}\vrule}\thinspace}

\title{An introduction to quantum gravity}

\author{Giampiero Esposito, \\
INFN, Sezione di Napoli, \\ 
Complesso Universitario di Monte S. Angelo, \\ 
Via Cintia, Edificio 6, 80126 Napoli, Italy}

\date{\today}
\begin{document}
\maketitle

\begin{abstract}
Quantum gravity was born as that branch of modern
theoretical physics that tries to unify its guiding principles,
i.e., quantum mechanics and general relativity. Nowadays it is
providing new insight into the unification of all fundamental
interactions, while giving rise to new developments in mathematics.  
The various competing theories, e.g. string theory and loop
quantum gravity, have still to be checked against observations.
We review the classical and quantum foundations necessary to study
field-theory approaches to quantum gravity, 
the passage from old to new unification in quantum
field theory, canonical quantum gravity, 
the use of functional integrals, the properties of 
gravitational instantons, the use of spectral zeta-functions 
in the quantum theory of the universe, Hawking radiation, some
theoretical achievements and some key experimental issues. 
\end{abstract}

\section{Introduction}

The aim of theoretical physics is to provide a clear conceptual framework
for the wide variety of natural phenomena, so that not only are we able to
make accurate predictions to be checked against observations, but the 
underlying mathematical structures of the world we live in can also become
sufficiently well understood by the scientific community. What are therefore
the key elements of a mathematical description of the physical world? Can
we derive all basic equations of theoretical physics from a set of symmetry
principles? What do they tell us about the origin and evolution of the
universe? Why is gravitation so peculiar with respect to all other 
fundamental interactions?

The above questions have received careful consideration over the last decades,
and have led, in particular, to several approaches to a theory aimed at
achieving a synthesis of quantum physics on the one hand, and general
relativity on the other hand. This remains, possibly, the most important
task of theoretical physics. In early work in the thirties, 
Rosenfeld \cite{Rose30a,Rose30b}
computed the gravitational self-energy of a photon in the lowest order
of perturbation theory, and obtained a quadratically divergent result. 
With hindsight, one can say that Rosenfeld's result implies merely
a renormalization of charge rather than a non-vanishing photon mass
\cite{DeWi09}. A few years after Rosenfeld's 
papers \cite{Rose30a,Rose30b}, Bronstein realized
that the limitation posed by general relativity on the mass density 
radically distinguishes the theory from quantum electrodynamics and would
ultimately lead to the need to reject Riemannian geometry and perhaps
also to reject our ordinary concepts of space and time
\cite{Bron36,Rove01}. 

Indeed, since the merging of quantum theory and special 
relativity has given rise to quantum field theory in Minkowski
spacetime, while quantum field theory and classical general relativity,
taken without modifications, have given rise to an incomplete 
scheme such as quantum field theory
in curved spacetime \cite{Full89}, which however predicts substantially 
novel features like Hawking radiation \cite{Hawk74,Hawk75}, 
here outlined in section 7, one is led
to ask what would result from the ``unification'' of quantum field
theory and gravitation, despite the lack of a quantum gravity phenomenology
in earth-based laboratories.
The resulting theory is expected to suffer from ultraviolet divergences
\cite{Wein79}, and the one-loop \cite{HoVe74} and two-loop
\cite{GoSa86} calculations for pure gravity are outstanding pieces of work. 
As is well described in Ref. \cite{Wein79},
if the coupling constant of a field theory has dimension 
${\rm mass}^{d}$ in ${\hbar}=c=1$ units, then the integral for a
Feynman diagram of order $N$ behaves at large momenta like
$\int p^{A-Nd}dp$, where $A$ depends on the physical process 
considered but not on the order $N$. Thus, the ``harmful'' interactions
are those having negative values of $d$, which is precisely the case
for Newton's constant $G$, where $d=-2$, since
$G=6.67 \times 10^{-39}{\rm GeV}^{-2}$ in ${\hbar}=c=1$ units. 
More precisely, since the scalar curvature 
contains second derivatives of the
metric, the corresponding momentum-space vertex functions behave
like $p^{2}$, and the propagator like $p^{-2}$. In $d$
dimensions each loop integral contributes $p^{d}$, so that with
$L$ loops, $V$ vertices and $P$ internal lines, the superficial
degree $D$ of divergence of a Feynman diagram is given 
by \cite{Espo94} 
\begin{equation}
D=dL+2V-2P.
\label{(3.4.11)}
\end{equation}
Moreover, a topological relation holds:
\begin{equation}
L=1-V+P,
\label{(3.4.12)}
\end{equation}
which leads to \cite{Espo94}
\begin{equation}
D=(d-2)L+2.
\label{(3.4.13)}
\end{equation}
In other words, $D$ increases with increasing loop order for
$d>2$, so that it clearly leads to a non-renormalizable theory.

A quantum theory of gravity is expected, for example, to shed new light 
on singularities in classical cosmology. More precisely, the singularity
theorems prove that the Einstein theory of general relativity leads to the
occurrence of spacetime singularities in a generic way \cite{HaPe70}.
At first sight one might be tempted to conclude that a breakdown of all
physical laws occurred in the past, or that general relativity is severely
incomplete, being unable to predict what came out of a singularity. It has
been therefore pointed out that all these pathological features result from 
the attempt of using the Einstein theory well beyond its limit of validity,
i.e. at energy scales where the fundamental theory is definitely more
involved. General relativity might be therefore viewed as a low-energy limit
of a richer theory, which achieves the synthesis of both the
{\bf basic principles} of modern physics and the 
{\bf fundamental interactions} in the form currently known.

So far, no less than 16 major approaches to quantum gravity have been
proposed in the literature. Some of them make a direct or indirect use
of the action functional to develop a Lagrangian or Hamiltonian
framework. They are as follows. 
\vskip 0.3cm
\noindent
1. Canonical quantum gravity 
\cite{Berg49,BeBr49,Dira58,Dira59,
DeWi67a,IsKa84a,IsKa84b,Asht86,EsGS95,Thie07}.
\vskip 0.3cm
\noindent
2. Manifestly covariant quantization \cite{Misn57,DeWi67b, 
HoVe74,GoSa86,Avra91,Vilk92,Canf05,Kief07}.
\vskip 0.3cm
\noindent
3. Euclidean quantum gravity \cite{GiHa77,Hawk78}.
\vskip 0.3cm
\noindent
4. R-squared gravity \cite{Stel77}.
\vskip 0.3cm
\noindent
5. Supergravity \cite{FrNF76,Nieu81}.
\vskip 0.3cm
\noindent
6. String and brane theory \cite{Witt02,Horo05,Barv06}. 
\vskip 0.3cm
\noindent
7. Renormalization group and Weinberg's asymptotic 
safety \cite{Reut98,LaRe02}.
\vskip 0.3cm
\noindent
8. Non-commutative geometry \cite{CoMa08,GrBo10}.
\vskip 0.3cm
\noindent
Among these 8 approaches, string theory is peculiar because it is
not field-theoretic, spacetime points being replaced by extended
structures such as strings.

A second set of approaches relies instead upon 
different mathematical structures
with a more substantial (but not complete) departure from
conventional pictures, i.e.
\vskip 0.3cm
\noindent
9. Twistor theory \cite{Penr72,Penr99}. 
\vskip 0.3cm
\noindent
10. Asymptotic quantization \cite{Gero71,Asht81}.
\vskip 0.3cm
\noindent
11. Lattice formulation \cite{MePe87,CaMa92}.
\vskip 0.3cm
\noindent
12. Loop space representation \cite{RoSm88,RoSm90,Rove04,Thie08,Vita11}.
\vskip 0.3cm
\noindent
13. Quantum topology \cite{Isha89}, motivated by Wheeler's
quantum geometrodynamics \cite{Whee57}.
\vskip 0.3cm
\noindent
14. Simplicial quantum gravity \cite{Gion98,ACGM99,Loll07,Ambj09}
and null-strut calculus \cite{KhLM89}.
\vskip 0.3cm
\noindent
15. Condensed-matter view: the universe in a helium droplet \cite{Volo03}.
\vskip 0.3cm
\noindent
16. Affine quantum gravity \cite{Klau06}.

After such a concise list of a broad range 
of ideas, we hereafter focus on the presentation
of some very basic properties which underlie whatever treatment of
classical and quantum gravity, and are therefore of interest for the
general reader rather than (just) the specialist. He or she should revert 
to the above list only after having gone through the material in sections
2--7.

\section{Classical and quantum foundations}

Before any attempt to quantize gravity we should spell out how
classical gravity can be described in modern language. This is done
in the subsection below.

\subsection{Lorentzian spacetime and gravity}

In modern physics, thanks to the work of Einstein \cite{Eins16},
space and time are unified into the spacetime manifold $(M,g)$, where
the metric $g$ is a real-valued symmetric bilinear map
$$
g:T_{p}(M) \times T_{p}(M) \rightarrow {\bf R}
$$
of Lorentzian signature. The latter feature gives rise to the light-cone
structure of spacetime, with vectors being divided into timelike, null or
spacelike depending on whether $g(X,X)$ is negative, vanishing or positive,
respectively. The classical laws of nature are written in tensor language,
and {\it gravity is the curvature of spacetime}. In the theory of general
relativity, gravity couples to the energy-momentum tensor of matter through
the Einstein equations
\begin{equation}
R_{\mu \nu}-{1\over 2}g_{\mu \nu}R={8 \pi G \over c^{4}}T_{\mu \nu}.
\label{(1)}
\end{equation}
The Einstein--Hilbert action functional for gravity, giving rise to
Eq. (\ref{(1)}), is diffeomorphism-invariant, and hence general relativity
belongs actually to the general set of theories ruled by an
infinite-dimensional \cite{DeWi65} invariance group
(or pseudo-group). With hindsight, following DeWitt \cite{DeWi05},
one can say that general relativity was actually the
first example of a non-Abelian gauge theory (about 38 years before
Yang--Mills theory \cite{YaMi54}).

Note that the spacetime manifold is actually an equivalence class
of pairs $(M,g)$, where two metrics are viewed as equivalent if
one can be obtained from the other through the action of the
diffeomorphism group ${\rm Diff}(M)$. The metric is an additional
geometric structure that does not necessarily solve any 
field equation.

\subsection{From Schr\"{o}dinger to Feynman}

Quantum mechanics deals instead, mainly, with a probabilistic description 
of the world on atomic or sub-atomic scale. It tells us that, on such
scales, the world can be described by a Hilbert space structure, or
suitable generalizations. Even in the relatively simple case of the
hydrogen atom, the appropriate Hilbert space is infinite-dimensional, but
finite-dimensional Hilbert spaces play a role as well. For example, the
space of spin-states of a spin-$s$ particle is 
${\bf C}^{2s+1}$ and is therefore
finite-dimensional. Various pictures or formulations of quantum mechanics
have been developed over the years, and their key elements can be summarized
as follows:
\renewcommand{\labelenumi}{\textup{(\roman{enumi})}}
\begin{enumerate}
\itemsep=0pt

\item In the {\it Schr\"{o}dinger picture}, one deals with wave functions
evolving in time according to a first-order equation. More precisely,
in an abstract Hilbert space ${\cal H}$, one studies the Schr\"{o}dinger
equation
\begin{equation}
{\rm i}{\hbar}{d \psi \over dt}={\hat H} \psi,
\label{(2)}
\end{equation}
where the state vector $\psi$ belongs to ${\cal H}$, while ${\hat H}$ is
the Hamiltonian operator. In wave mechanics, the emphasis is more
immediately put on partial differential equations, with the wave function
viewed as a complex-valued map $\psi: (x,t) 
\rightarrow {\bf C}$ obeying
the equation
\begin{equation}
{\rm i}{\hbar}{\partial \psi \over \partial t}
=\left(-{{\hbar}^{2}\over 2m}\bigtriangleup + V \right) \psi,
\label{(3)}
\end{equation}
where $- \bigtriangleup$ is the Laplacian in Cartesian coordinates on
${\bf R}^{3}$ (with this sign convention, its 
symbol is positive-definite).
\item In the {\it Heisenberg picture}, what evolves in time are instead
the operators, according to the first-order equation
\begin{equation} 
{\rm i}{\hbar}{d{\hat A}\over dt}=[{\hat A},{\hat H}].
\label{(4)}
\end{equation}
Heisenberg performed a quantum mechanical re-interpretation of kinematic
and mechanical relations \cite{Heis25}
because he wanted to formulate quantum theory
in terms of observables only.

\item  In the {\it Dirac quantization}, from an assessment of the
Heisenberg approach and of Poisson brackets \cite{Dira25},
one discovers that quantum
mechanics can be made to rely upon the basic commutation relations
involving position and momentum operators:
\begin{equation}
[{\hat q}^{j},{\hat q}^{k}]=[{\hat p}_{j},{\hat p}_{k}]=0,
\label{(5)}
\end{equation}
\begin{equation}
[{\hat q}^{j},{\hat p}_{k}]={\rm i}{\hbar}\delta_{\; k}^{j}.
\label{(6)}
\end{equation}
For generic operators depending on ${\hat q},{\hat p}$ variables,
their formal Taylor series, jointly with application of 
(\ref{(5)}) and (\ref{(6)}), should yield their commutator.
\item  {\it Weyl quantization}. The operators satisfying the canonical
commutation relations (\ref{(6)}) cannot be both 
bounded \cite{Espo04}, whereas it would be nice
to have quantization rules not involving unbounded operators and domain
problems. For this purpose, one can consider the strongly continuous
one-parameter unitary groups having position and momentum as their
infinitesimal generators. These read as 
$V(t) \equiv {\rm e}^{{\rm i}t{\hat q}}$, 
$U(s) \equiv {\rm e}^{{\rm i}s{\hat p}}$, 
and satisfy the Weyl form of canonical
commutation relations, which is given by
\begin{equation}
U(s)V(t)={\rm e}^{{\rm i}st{\hbar}}V(t)U(s).
\label{(7)}
\end{equation}
Here the emphasis was, for the first time, on group-theoretical methods,
with a substantial departure from the historical development, that relied
instead heavily on quantum commutators and their relation with
classical Poisson brackets.
\item {\it Feynman quantization} (i.e., Lagrangian approach).
The Weyl approach is very elegant and far-sighted,
with several modern applications
\cite{Espo04}, but still has to do with a more rigorous
way of doing canonical quantization, which is not suitable for an
inclusion of relativity. A spacetime approach to ordinary quantum mechanics
was instead devised by Feynman \cite{Feyn48}
(and partly Dirac himself \cite{Dira33}), who proposed
to express the Green kernel of the Schr\"{o}dinger equation in the form
\begin{equation}
G[x_{f},t_{f};x_{i},t_{i}]=\int_{{\rm all \; paths}}
{\rm e}^{{\rm i}S / {\hbar}} d\mu,
\label{(8)}
\end{equation}
where $d\mu$ is a suitable (putative) measure on the set of all spacetime
paths (including continuous, piecewise continuous, or even discontinuous
paths) matching the initial and final conditions. This point of view has
enormous potentialities in the quantization of field theories, since it
preserves manifest covariance and the full symmetry group, being derived
from a Lagrangian.
\end{enumerate}

It should be stressed that quantum mechanics regards wave functions
only as a technical tool to study bound states (corresponding to the
discrete spectrum of the Hamiltonian operator ${\hat H}$),
scattering states (corresponding instead to the continuous spectrum
of ${\hat H}$), and to evaluate probabilities (of finding the values
taken by the observables of the theory). Moreover, it is meaningless to
talk about an elementary phenomenon on atomic (or sub-atomic) scale
unless it is registered \cite{WhZu83}, 
and quantum mechanics in the laboratory
needs also an external observer and assumes the so-called reduction
of the wave packet (see \cite{Espo04} and references therein).
There exist indeed different interpretations of quantum mechanics,
e.g. Copenhagen \cite{WhZu83}, hidden variables \cite{Bell66},
many worlds \cite{Ever57,DeWi71}.

\subsection{Spacetime singularities}

Now we revert to the geometric side. In Riemannian or pseudo-Riemannian
geometry, geodesics are curves whose tangent vector $X$ moves by
parallel transport \cite{HaEl73}, so that eventually
\begin{equation}
{dX^{\lambda}\over ds}+\Gamma_{\; \mu \nu}^{\lambda}
X^{\mu}X^{\nu}=0,
\label{(9)}
\end{equation}
where $s$ is the affine parameter and 
$\Gamma_{\; \mu \nu}^{\lambda}$ are the
connection coefficients. In general relativity, timelike geodesics
correspond to the trajectories of freely moving observers, while
null geodesics describe the trajectories of zero-rest-mass particles
(section 8.1 of Ref. \cite{HaEl73}). Moreover, a
spacetime $(M,g)$ is said to be singularity-free if {\it all timelike
and null geodesics can be extended to arbitrary values of their
affine parameter}. At a spacetime singularity in general relativity,
all laws of classical physics would break down, because one would witness
very pathological events such as the sudden disappearance of freely
moving observers, and one would be completely unable to predict what
came out of the singularity.
In the sixties, Penrose \cite{Penr65}
proved first an important theorem on the
occurrence of singularities in gravitational collapse (e.g. formation
of black holes). Subsequent work by Hawking
\cite{Hawk65, Hawk66a, Hawk66b, Hawk66c,Hawk67},
Geroch \cite{Gero66}, Ellis and Hawking \cite{HaEl65, ElHa68},
Hawking and Penrose \cite{HaPe70}
proved that spacetime singularities are generic
properties of general relativity, provided that physically realistic
energy conditions hold. Very little analytic use of the Einstein
equations is made, whereas the key role emerges of topological and
global methods in general relativity.

On the side of singularity theory in classical cosmology, explicit 
mention should be made of the work in Ref. \cite{Beli70}, since
it has led to significant progress by Damour et al. \cite{Damo03},
despite having failed to prove singularity avoidance in classical
cosmology. As pointed out in Ref. \cite{Damo03}, the work by Belinsky 
et al. is remarkable because it gives a description of the generic
asymptotic behaviour of the gravitational field in four-dimensional
spacetime in the vicinity of a spacelike singularity. Interestingly,
near the singularity the spatial points essentially decouple, i.e.
the evolution of the spatial metric at each spatial point is 
asymptotically governed by a set of second-order, non-linear ordinary
differential equations in the time variable \cite{Beli70}.
Moreover, the use of qualitative Hamiltonian methods leads naturally
to a billiard description of the asymptotic evolution, where the 
logarithms of spatial scale factors define a geodesic motion in a
region of the Lobachevskii plane, interrupted by geometric reflections
against the walls bounding this region. Chaos follows because the
Bianchi IX billiard has finite volume \cite{Damo03}. A self-contained
derivation of the billiard picture for inhomogeneous solutions in
$D$ dimensions, with dilaton and $p$-form gauge fields, has been
obtained in Ref. \cite{Damo03}.  

\subsection{Unification of all fundamental interactions}

The fully established unifications of modern physics are as follows.
\begin{enumerate}
\itemsep=0pt

\item {\it Maxwell}: electricity and magnetism are unified into
electromagnetism. All related phenomena can be described by an
antisymmetric rank-two tensor field, and derived from a one-form,
called the potential.
\item {\it Einstein}: space and time are unified into the spacetime
manifold. Moreover, inertial and gravitational mass, conceptually different,
are actually unified as well.
\item {\it Standard model of particle physics}: electromagnetic, weak
and strong forces are unified by a non-Abelian gauge theory, normally
considered in Minkowski spacetime (this being the base space in
fibre-bundle language).
\end{enumerate}

The physics community is now familiar with a picture relying upon four
fundamental interactions: electromagnetic, weak, strong and gravitational.
The large-scale structure of the universe, however, is ruled by gravity
only. All unifications beyond Maxwell involve non-Abelian gauge groups
(either Yang--Mills or Diffeomorphism group). At least three 
extreme views have been developed along the years, i.e.,
\begin{enumerate}
\itemsep=0pt

\item  Gravity arose first, temporally,
in the very early Universe, then all other fundamental interactions.
\item  Gravity might result from Quantum Field Theory (this was the
Sakharov idea \cite{Sakh68}).
\item  The vacuum of particle physics is regarded as a cold quantum liquid
in equilibrium. Protons, gravitons and gluons are viewed as collective
excitations of this liquid \cite{Volo03}.
\end{enumerate}

\section{Canonical quantum gravity}

Although Hamiltonian methods differ substantially from the
Lagrangian approach used in the construction of the functional
integral (see the following sections), 
they remain nevertheless of great importance both
in cosmology and in light of modern developments in
canonical quantum gravity \cite{Asht86,Rove04,Thie07}, 
which is here presented within the original 
framework of quantum geometrodynamics.
For this purpose, it may be useful to describe the main
ideas of the Arnowitt-Deser-Misner (hereafter referred to
as ADM) formalism. This is a canonical formalism for
general relativity that enables one to re-write Einstein's
field equations in first-order form and explicitly solved
with respect to a time variable. For this purpose, one
assumes that four-dimensional spacetime $(M,g)$ can be
foliated by a family of $t={\rm constant}$ spacelike surfaces
$S_{t}$, giving rise to a $3+1$ decomposition of the original
4-geometry. The basic geometric data of this decomposition
are as follows \cite{Espo94}.

(1) The induced 3-metric $h$ of the three-dimensional
spacelike surfaces $S_{t}$. This yields the intrinsic geometry
of the three-space. $h$ is also called the first fundamental
form of $S_{t}$, and is positive-definite with our
conventions.

(2) The way each $S_{t}$ is imbedded in $(M,g)$. This is
known once we are able to compute the spatial part of the
covariant derivative of the normal $n$ to $S_{t}$. On denoting
by $\nabla$ the four-connection of $(M,g)$, one is thus led to
define the tensor
\begin{equation}
K_{ij} \equiv - \nabla_{j}n_{i}.
\label{(2.4.1)}
\end{equation}
Note that $K_{ij}$ is symmetric if and only if $\nabla$ is
symmetric. In general relativity, an equivalent
definition of $K_{ij}$ is $K_{ij} \equiv -{1\over 2}
(L_{n}h)_{ij}$, where $L_{n}$ denotes the Lie derivative 
along the normal to $S_{t}$. The tensor $K$ is called
extrinsic-curvature tensor, or second fundamental form of
$S_{t}$.

(3) How the coordinates are propagated off the surface 
$S_{t}$. For this purpose one defines the vector 
$(N,N^{1},N^{2},N^{3})dt$ connecting the point $(t,x^{i})$
with the point $(t+dt,x^{i})$. Thus, given the surface
$x^{0}=t$ and the surface $x^{0}=t+dt$, $Ndt \equiv d\tau$
specifies a displacement normal to the surface
$x^{0}=t$. Moreover, $N^{i}dt$ yields the displacement from
the point $(t,x^{i})$ to the foot of the normal to $x^{0}=t$
through $(t+dt,x^{i})$. In other words, the 
$N^{i}$ arise since the $x^{i}={\rm constant}$ lines do not coincide
in general with the normals to the $t={\rm constant}$ surfaces.
According to a well-established terminology, $N$ is the lapse
function, and the $N^{i}$ are the shift functions. They are
the tool needed to achieve the desired space-time foliation.

In light of points (1)--(3) as above, the 4-metric $g$ can
be locally cast in the form
\begin{equation}
g=h_{ij}\Bigr(dx^{i}+N^{i}dt\Bigr) \otimes
\Bigr(dx^{j}+N^{j}dt\Bigr)-N^{2}dt \otimes dt .
\label{(2.4.2)}
\end{equation}
This implies that 
\begin{equation}
g_{00}=-\Bigr(N^{2}-N_{i}N^{i}\Bigr),
\label{(2.4.3)}
\end{equation}
\begin{equation}
g_{i0}=g_{0i}=N_{i},
\label{(2.4.4)}
\end{equation}
\begin{equation}
g_{ij}=h_{ij},
\label{(2.4.5)}
\end{equation}
whereas, using the property $g^{\lambda \nu}g_{\nu \mu}=
\delta_{\; \; \mu}^{\lambda}$, one finds 
\begin{equation}
g^{00}=-{1\over N^{2}},
\label{(2.4.6)}
\end{equation}
\begin{equation}
g^{i0}=g^{0i}={N^{i}\over N^{2}},
\label{(2.4.7)}
\end{equation}
\begin{equation}
g^{ij}=h^{ij}-{N^{i}N^{j}\over N^{2}}.
\label{(2.4.8)}
\end{equation}
Interestingly, the covariant $g_{ij}$ and $h_{ij}$
coincide, whereas the contravariant 
$g^{ij}$ and $h^{ij}$ differ as
shown in (\ref{(2.4.8)}). In terms of $N,N^{i}$ and $h$, the
extrinsic-curvature tensor defined in (\ref{(2.4.1)}) takes the form
\begin{equation}
K_{ij} \equiv {1\over 2N} \biggr(-{\partial h_{ij}
\over \partial t}+N_{i \mid j} +N_{j \mid i}\biggr),
\label{(2.4.9)}
\end{equation}
where the stroke $\mid$ denotes covariant differentiation
on the spacelike 3-surface $S_{t}$, and indices of $K_{ij}$
are raised using $h^{il}$. Equation (\ref{(2.4.9)}) can be also 
written as 
\begin{equation}
{\partial h_{ij} \over \partial t}=
N_{i \mid j} +N_{j \mid i} -2NK_{ij}.
\label{(2.4.10)}
\end{equation}
Equation (\ref{(2.4.10)}) should be supplemented by another first-order
equation expressing the time evolution of 
$K_{ij}$ (recall that $\pi^{ij}$ is related to $K^{ij}$), i.e.
\begin{eqnarray}
{\partial K_{ij}\over \partial t}&=& -N_{\mid ij}
+N \Bigr[{ }^{(3)}R_{ij}+K_{ij}({\rm tr}K)-2K_{im}K_{j}^{\; m}
\Bigr] \nonumber \\
&+& \Bigr[N^{m}K_{ij \mid m}+N_{\mid i}^{m}K_{jm}
+N_{\mid j}^{m}K_{im}\Bigr]. 
\end{eqnarray}

On using the ADM variables described so far, the form of the
action integral $I$ for pure gravity that is stationary
under variations of the metric vanishing on the boundary
is (in $c=1$ units) 
\begin{eqnarray}
I& \equiv & {1\over 16 \pi G}\int_{M}{ }^{(4)}R 
\; \sqrt{-g} \; d^{4}x
+{1\over 8 \pi G}\int_{\partial M}
K_{i}^{i} \; \sqrt{h} \; d^{3}x \nonumber \\
&=& {1\over 16 \pi G} \int_{M}\biggr[{ }^{(3)}R +K_{ij}K^{ij}-
{\Bigr(K_{i}^{i}\Bigr)}^{2} \biggr]
N\sqrt{h} \; d^{3}x \; dt.
\label{(2.4.11)}
\end{eqnarray}
The boundary term appearing in (\ref{(2.4.11)}) is necessary since 
${ }^{(4)}R$ contains second derivatives of the metric,
and integration by parts in the Einstein--Hilbert part
$$
I_{H} \equiv {1\over 16 \pi G} \int_{M}
{ }^{(4)}R \; \sqrt{-g} \; d^{4}x
$$ 
of the action also leads
to a boundary term equal to $-{1\over 8 \pi G}
\int_{\partial M}K_{i}^{i} \; \sqrt{h} \; d^{3}x$. On denoting
by $G_{\mu \nu}$ the Einstein tensor
$G_{\mu \nu} \equiv { }^{(4)}R_{\mu \nu}-{1\over 2}g_{\mu \nu} \;
{ }^{(4)}R$, and defining 
\begin{equation}
\delta \Gamma_{\mu \nu}^{\rho} \equiv
{1\over 2}g^{\rho \lambda}
\left[\nabla_{\mu} \Bigr(\delta g_{\lambda \nu}\Bigr)
+\nabla_{\nu} \Bigr(\delta g_{\lambda \mu}\Bigr)
-\nabla_{\lambda} \Bigr(\delta g_{\mu \nu}\Bigr) \right],
\label{(2.4.12)}
\end{equation}
one then finds \cite{York86}
\begin{equation}
(16\pi G) \delta I_{H} =-\int_{M}\sqrt{-g} \; G^{\mu \nu} \;
\delta g_{\mu \nu} \; d^{4}x
+ \int_{\partial M}\sqrt{-g}
\Bigr(g^{\mu \nu}\delta_{\rho}^{\sigma}
-g^{\mu \sigma}\delta_{\rho}^{\nu}\Bigr)
\delta \Gamma_{\mu \nu}^{\rho}
\Bigr(d^{3}x\Bigr)_{\sigma},
\label{(2.4.13)}
\end{equation}
which clearly shows that $I_{H}$ is stationary if the
Einstein equations hold, and the normal derivatives of
the variations of the metric vanish on the boundary
$\partial M$. In other words, $I_{H}$ is not stationary
under {\it arbitrary} variations of the metric, and 
stationarity is only achieved after adding to $I_{H}$
the boundary term appearing in (\ref{(2.4.11)}), 
if $\delta g_{\mu \nu}$ is set to zero on 
$\partial M$. Other useful forms
of the boundary term can be found in \cite{GiHa77,York86}. 
Note also that, strictly, in writing
down (\ref{(2.4.11)}) one should also take into account a term
arising from $I_{H}$ \cite{DeWi67a}:
\begin{equation}
I_{t} \equiv {1\over 8 \pi G} \int dt \int_{\partial M} 
d^{3}x \; \partial_{i}
\Bigr[\sqrt{h}\Bigr(K_{l}^{l} \; N^{i}
- h^{ij} \; N_{\mid j} \Bigr)\Bigr].
\end{equation}
However, we have not explicitly included $I_{t}$ since it
does not modify the results derived or described hereafter.

We are now ready to apply Dirac's technique to 
the Hamiltonian quantization of general relativity. 
This requires that all classical constraints which are first-class
are turned into operators that annihilate the wave functional
\cite{Espo94}. Hereafter, we assume that 
this step has already been performed.
As we know, consistency of the quantum constraints is proved if one
can show that their commutators lead to no new constraints \cite{Espo94}.
For this purpose, it may be useful to recall the equal-time
commutation relations of the canonical variables, i.e.
\begin{equation}
\Bigr[N(x),\pi(x')\Bigr]={\rm i} \delta(x,x') ,
\label{(2.4.14)}
\end{equation}
\begin{equation}
\Bigr[N_{j}(x),\pi^{k}(x')\Bigr]={\rm i} \delta_{j}^{k'},
\label{(2.4.15)}
\end{equation}
\begin{equation}
\Bigr[h_{jk},\pi^{l'm'}\Bigr]={\rm i} \delta_{jk}^{\; \; \; l'm'}.
\label{(2.4.16)}
\end{equation}
Note that, following \cite{DeWi67a}, primes have been used,
either on indices or on the variables themselves, to
distinguish different points of three-space. In other words,
one defines 
\begin{equation}
\delta_{i}^{j'} \equiv \delta_{i}^{j} \; \delta(x,x'),
\label{(2.4.17)}
\end{equation}
\begin{equation}
\delta_{ij}^{\; \; \; k'l'} \equiv \delta_{ij}^{\; \; \; kl}
\; \delta(x,x') ,
\label{(2.4.18)}
\end{equation}
\begin{equation}
\delta_{ij}^{\; \; \; kl} \equiv {1\over 2}
\Bigr(\delta_{i}^{k}\delta_{j}^{l}+\delta_{i}^{l}\delta_{j}^{k}
\Bigr) .
\label{(2.4.19)}
\end{equation}
The reader can check that, since \cite{DeWi67a}
\begin{equation}
{\cal H} \equiv \sqrt{h}\Bigr(K_{ij}K^{ij}-K^{2}
-{ }^{(3)}R \Bigr),
\label{(1.1.4)}
\end{equation}
\begin{equation}
{\cal H}^{i} \equiv -2 \pi_{,j}^{ij}-h^{il}\Bigr(2h_{jl,k}
-h_{jk,l}\Bigr)\pi^{jk},
\label{(1.1.5)}
\end{equation}
one has
\begin{eqnarray}
\Bigr[\pi(x),\pi^{i}(x')\Bigr]&=&\Bigr[\pi(x),{\cal H}^{i}(x')
\Bigr]=\Bigr[\pi(x),{\cal H}(x')\Bigr]=
\Bigr[\pi^{i}(x),{\cal H}^{j}(x')\Bigr] \nonumber\\
&=&\Bigr[\pi^{i}(x),{\cal H}(x')\Bigr]=0 .
\label{(2.4.20)}
\end{eqnarray}
It now remains to compute the three commutators
$\Bigr[{\cal H}_{i},{\cal H}_{j'}\Bigr]$,
$\Bigr[{\cal H}_{i},{\cal H}' \Bigr]$,
$\Bigr[{\cal H},{\cal H}' \Bigr]$. 
The first two commutators are obtained by using Eq. (\ref{(1.1.5)}) 
and defining ${\cal H}_{i} \equiv h_{ij}{\cal H}^{j}$.
Interestingly, ${\cal H}_{i}$ is homogeneous bilinear
in the $h_{ij}$ and $\pi^{ij}$, with the momenta always
to the right. As we said before,
following Dirac, the operator version of
constraints should annihilate the wave function since
the classical constraints are first-class (i.e. their Poisson
brackets are linear combinations of the constraints themselves).
This condition reads as 
\begin{equation}
\int_{S_{t}} {\cal H} \xi \; d^{3}x \; \psi =0 \; \; \; \;
\forall \xi ,
\label{(2.4.21)}
\end{equation}
\begin{equation}
\int_{S_{t}} {\cal H}_{i}\xi^{i} \; d^{3}x \; \psi =0 \; \; \; \; 
\forall \xi^{i} .
\label{(2.4.22)}
\end{equation}
In the applications to cosmology, Eq. (\ref{(2.4.21)}) is known as
the Wheeler--DeWitt equation, and the functional $\psi$ is then
called the {\it wave function of the universe} \cite{HaHa83}.
 
We begin by computing \cite{DeWi67a}
\begin{equation}
\left[h_{jk},{\rm i} \int_{S_{t}} {\cal H}_{k'} \delta \xi^{k'} \;
d^{3}x' \right]=
-h_{jk,l} \; \delta \xi^{l}-h_{lk} \; \delta \xi_{\; \; ,j}^{l}
-h_{jl} \; \delta \xi_{\; \; ,k}^{l},
\label{(2.4.23)}
\end{equation}
\begin{equation}
\left[\pi^{jk},{\rm i} \int_{S_{t}} {\cal H}_{k'}\delta \xi^{k'} \;
d^{3}x' \right]=
-\Bigr(\pi^{jk}\delta \xi^{l}\Bigr)_{,l}
+\pi^{lk} \; \delta \xi_{\; \; ,l}^{j}
+\pi^{jl} \; \delta \xi_{\; \; ,l}^{k}.
\label{(2.4.24)}
\end{equation}
This calculation shows that the ${\cal H}_{i}$ are
generators of three-dimensional coordinate transformations
${\overline x}^{i}=x^{i}+\delta \xi^{i}$.
Thus, by using the definition of structure constants
of the general coordinate-transformation group \cite{DeWi67a}, i.e.
\begin{equation}
c_{\; \; \; ij'}^{k''} \equiv
\delta_{\; \; \; i,l''}^{k''} \; \delta_{j'}^{l''}
-\delta_{\; \; \; j',l''}^{k''} \; \delta_{i}^{l''},
\label{(2.4.25)}
\end{equation}
the results (\ref{(2.4.23)})--(\ref{(2.4.24)}) may be used to show that 
\begin{equation}
\Bigr[{\cal H}_{j}(x),{\cal H}_{k}(x')\Bigr]=-{\rm i}
\int_{S_{t}} {\cal H}_{l''} \; c_{\; \; \; jk'}^{l''} \; d^{3}x'',
\label{(2.4.26)}
\end{equation}
\begin{equation}
\Bigr[{\cal H}_{j}(x),{\cal H}(x')\Bigr]=
{\rm i}{\cal H} \; \delta_{,j}(x,x') .
\label{(2.4.27)}
\end{equation}
Note that the only term of ${\cal H}$ which might lead to
difficulties is the one quadratic in the momenta. However,
all factors appearing in this term have homogeneous linear 
transformation laws under the three-dimensional 
coordinate-transformation group. They thus remain undisturbed in
position when commuted with ${\cal H}_{j}$ \cite{DeWi67a}.

Last, we have to study the commutator
$\Bigr[{\cal H}(x),{\cal H}(x')\Bigr]$. The following
remarks are in order:

(i) Terms quadratic in momenta contain no derivatives of
$h_{ij}$ or $\pi^{ij}$ with respect to three-space coordinates.
Hence they commute;

(ii) The terms $\sqrt{h(x)}\Bigr({ }^{(3)}R(x)\Bigr)$ and
$\sqrt{h(x')}\Bigr({ }^{(3)}R(x')\Bigr)$ contain no momenta,
so that they also commute;

(iii) The only commutators we are left with are the 
cross-commutators, and they can be evaluated by using the 
variational formula \cite{DeWi67a} 
\begin{eqnarray}
\delta \Bigr(\sqrt{h} \; { }^{(3)}R \Bigr) &=&
\sqrt{h} \; h^{ij}h^{kl}
\Bigr(\delta h_{ik,jl}-\delta h_{ij,kl}\Bigr) \nonumber \\
&-& \sqrt{h} \left[{ }^{(3)}R^{ij} -{1\over 2}
h^{ij} \Bigr({ }^{(3)}R \Bigr)\right] \delta h_{ij},
\label{(2.4.28)}
\end{eqnarray}
which leads to 
\begin{equation}
\left[\int_{S_{t}} {\cal H} \; \xi_{1} \; d^{3}x,
\int_{S_{t}} {\cal H} \; \xi_{2} \; d^{3}x \right] = 
{\rm i} \int_{S_{t}} {\cal H}^{l} \Bigr(\xi_{1} \; \xi_{2,l}-
\xi_{1,l} \; \xi_{2} \Bigr) \; d^{3}x .
\label{(2.4.29)}
\end{equation}
The commutators (\ref{(2.4.26)})--(\ref{(2.4.27)}) 
and (\ref{(2.4.29)}) clearly show that
the constraint equations of canonical quantum gravity are
first-class. The Wheeler-DeWitt equation 
(\ref{(2.4.21)}) is an equation on the superspace 
(here $\Sigma$ is a Riemannian 3-manifold diffeomorphic to $S_{t}$)
$$
S(\Sigma) \equiv 
Riem(\Sigma)/Diff(\Sigma).
$$ 
In this quotient space, two Riemannian metrics on $\Sigma$  
are identified if they are related through the action of the
diffeomorphism group ${\rm Diff}(\Sigma)$.

Two very useful classical formulae frequently 
used in Lorentzian canonical gravity are 
\begin{equation}
{\cal H} \equiv (16\pi G)G_{ijkl}p^{ij}p^{kl}
-{\sqrt{h}\over 16\pi G} \Bigr({ }^{(3)}R\Bigr) ,
\label{(2.4.30)}
\end{equation}
\begin{equation}
{\cal H} \equiv (16\pi G)^{-1} \Bigr[G^{ijml}K_{ij}K_{ml}
-\sqrt{h} \Bigr( { }^{(3)}R \Bigr) \Bigr],
\label{(2.4.31)}
\end{equation}
where the rank-4 tensor density
is the DeWitt supermetric on superspace,
with covariant and contravariant forms 
\begin{equation}
G_{ijkl} \equiv {1\over 2 \sqrt{h}}
\Bigr(h_{ik}h_{jl}+h_{il}h_{jk}-h_{ij}h_{kl}\Bigr),
\label{(2.4.32)}
\end{equation}
\begin{equation}
G^{ijkl} \equiv {\sqrt{h}\over 2}
\Bigr(h^{ik}h^{jl}+h^{il}h^{jk}-2h^{ij}h^{kl}\Bigr),
\label{(2.4.33)}
\end{equation}
and $p^{ij}$ is here defined as $-{\sqrt{h}\over 16 \pi G}
\Bigr(K^{ij}-h^{ij}K \Bigr)$. Note that the factor $-2$
multiplying $h^{ij}h^{kl}$ in (\ref{(2.4.33)}) is needed so as
to obtain the identity 
\begin{equation}
G_{ijmn}G^{mnkl}={1\over 2}
\Bigr(\delta_{i}^{k}\delta_{j}^{l}+
\delta_{i}^{l}\delta_{j}^{k} \Bigr).
\label{(2.4.34)}
\end{equation}
Equation (\ref{(2.4.30)}) clearly shows that 
${\cal H}$ contains a part
quadratic in the momenta and a part proportional to 
${ }^{(3)}R$ (cf. (\ref{(1.1.4)})). On quantization, it is then hard
to give a well-defined meaning to the second functional
derivative ${\delta^{2}\over \delta h_{ij} \delta h_{kl}}$,
whereas the occurrence of ${ }^{(3)}R$ makes it even 
more difficult to solve exactly the Wheeler-DeWitt equation.

It should be stressed that wave functions built from the functional
integral (see the following sections) which generalizes the
path integral of ordinary quantum mechanics (see (\ref{(8)}))
do not solve the Wheeler--DeWitt equation (\ref{(2.4.21)}) 
unless some suitable assumptions are made \cite{HaHa91}, and
counterexamples have been built, i.e. a functional integral for the
wave function of the universe which does not solve the
Wheeler--DeWitt equation \cite{DeWi99}.

\section{From old to new unification}

Here we outline how the space-of-histories formulation provides a 
common ground for describing the `old' and `new' unifications of 
fundamental theories.

\subsection{Old unification}

Quantum field theory begins once an action functional $S$ is given, since
{\it the first and most fundamental assumption of 
quantum theory is that every isolated dynamical system 
can be described by a characteristic action functional} 
\cite{DeWi65}.  The Feynman approach makes it necessary to
consider an infinite-dimensional manifold such as the space $\Phi$ of all
field histories $\varphi^{i}$.
On this space there exist (in the case of gauge theories) vector fields
\begin{equation}
Q_{\alpha}=Q_{\; \alpha}^{i} {\delta \over \delta \varphi^{i}}
\label{(10)}
\end{equation}
that leave the action invariant, i.e. \cite{DeWi05}
\begin{equation}
Q_{\alpha}S=0.
\label{(11)}
\end{equation}
The Lie brackets of these vector fields lead to a classification
of all gauge theories known so far.

\subsection{Type-I gauge theories}

The peculiar property of type-I gauge theories is that these Lie
brackets are equal to linear combinations of the vector fields themselves,
with structure constants, i.e. \cite{DeWi03}
\begin{equation}
[Q_{\alpha},Q_{\beta}]=C_{\; \alpha \beta}^{\gamma} \; Q_{\gamma},
\label{(12)}
\end{equation}
where ${\delta C_{\; \alpha \beta}^{\gamma}
\over \delta \varphi^{i}}=0$. 
The Maxwell, Yang--Mills,
Einstein theories are all example of type-I theories (this is the
`unifying feature'). All of them, being gauge theories, need
supplementary conditions, since the second functional derivative of
$S$ is not an invertible operator. After imposing such conditions, the
theories are ruled by an invertible 
differential operator of D'Alembert type (or
Laplace type, if one deals instead with Euclidean field theory), or a
non-minimal operator at the very worst (for arbitrary choices of
gauge parameters). For example, when Maxwell theory is quantized via
functional integrals in the Lorenz \cite{Lore67} gauge, one deals
with a gauge-fixing functional
\begin{equation}
\Phi(A)=\nabla^{\mu}A_{\mu},
\label{(13)}
\end{equation}
and the second-order differential operator acting on the potential
in the gauge-fixed action functional reads as
\begin{equation}
P_{\mu}^{\; \nu}=-\delta_{\mu}^{\; \nu}\cstok{\ }+R_{\mu}^{\; \; \nu}
+\left(1-{1\over \alpha}\right)\nabla_{\mu}\nabla^{\nu},
\label{(14)}
\end{equation}
where $\alpha$ is an arbitrary gauge parameter. The Feynman choice
$\alpha=1$ leads to the minimal operator
$$
{\widetilde P}_{\mu}^{\; \nu}=-\delta_{\mu}^{\; \nu}\cstok{\ }
+R_{\mu}^{\; \nu},
$$
which is the standard wave operator on vectors in curved spacetime.
Such operators play a leading role in the one-loop expansion of the
Euclidean effective action, i.e. the quadratic order in $\hbar$ in
the asymptotic expansion of the functional ruling the quantum theory
with positive-definite metrics.

The closure property expressed by Eq. (\ref{(12)}) implies that the
gauge group decomposes the space of histories $\Phi$ into sub-spaces
to which the $Q_{\alpha}$ are tangent. These sub-spaces are known
as orbits, and $\Phi$ may be viewed as a principal fibre bundle of 
which the orbits are the fibres. The space of orbits is, strictly,
the quotient space $\Phi / {\cal G}$, where ${\cal G}$ is the proper
gauge group, i.e. the set of transformations of $\Phi$ into itself 
obtained by exponentiating the infinitesimal gauge transformation
\begin{equation}
\delta \varphi^{i}=Q_{\alpha}^{i} \; \delta \xi^{\alpha},
\end{equation}
and taking products of the resulting exponential maps.
Suppose one performs the transformation \cite{DeWi05}
\begin{equation}
\varphi^{i} \rightarrow I^{A},K^{\alpha}
\end{equation}
from the field variables $\varphi^{i}$ to a set of fibre-adapted
coordinates $I^{A}$ and $K^{\alpha}$. With this notation, the 
$I$'s label the fibres, i.e. the points in $\Phi / {\cal G}$, and
are gauge invariant because
\begin{equation}
Q_{\alpha}I^{A}=0.
\end{equation}
The $K$'s label the points within each fibre, and one often makes
specific choices for the $K$'s, corresponding to the choice of
supplementary condition \cite{DeWi65}, more frequently called
gauge condition. One normally picks out a base point $\varphi_{*}$
in $\Phi$ and chooses the $K's$ to be local functionals of the 
$\varphi$'s in such a way that the formula
\begin{equation}
{\widehat {\cal F}}_{\beta}^{\alpha} \equiv
Q_{\beta}K^{\alpha}=K_{,i}^{\alpha} \; Q_{\beta}^{i}
\end{equation}
defines a non-singular differential operator,
called the {\it ghost operator}, at and in a 
neighbourhood of $\varphi_{*}$. Thus, what is often called choosing
a gauge amounts to choosing a hypersurface $K^{\alpha}={\rm constant}$
in a fibre-adapted coordinate patch. The fields acted upon by the ghost
operator are called {\it ghost fields}, and have opposite statistics 
with respect to the fields occurring in the gauge-invariant action
functional (see Refs. \cite{Feyn63,FaPo67,DeWi67b} for the first time
that ghost fields were considered in quantized gauge theories).
The gauge-fixed action in the functional integral reads as \cite{DeWi05}
\begin{equation}
S_{{\rm g.f.}}=S+{1\over 2}
K^{\alpha}\omega_{\alpha \beta'}K^{\beta'},
\end{equation}
where $\omega_{\alpha \beta'}$ is a non-singular matrix of gauge
parameters (strictly, it is written with matrix notation, but
it contains Dirac's delta, i.e. $\omega_{\alpha \beta'}
\equiv \omega_{\alpha \beta}\delta(x,x')$).

\subsection{Type-II gauge theories}

For type-II gauge theories, Lie brackets of vector fields $Q_{\alpha}$ are
as in Eq. (\ref{(12)}) for type-I theories, 
but the structure constants are promoted
to structure functions. An example is given by simple supergravity (a
supersymmetric \cite{GoLi71,WeZu74} gauge theory of gravity, with a symmetry
relating bosonic and fermionic fields) in four spacetime dimensions, with
auxiliary fields \cite{Nieu81}.

\subsection{Type-III gauge theories}

In this case, the Lie bracket (\ref{(12)}) is generalized by
\begin{equation}
[Q_{\alpha},Q_{\beta}]=C_{\; \alpha \beta}^{\gamma} \; Q_{\gamma}
+U_{\; \alpha \beta}^{i} \; S_{,i},
\label{(15)}
\end{equation}
and it therefore reduces to (\ref{(12)}) 
only on the {\it mass-shell}, i.e.
for those field configurations satisfying the Euler--Lagrange equations.
An example is given by theories with gravitons and gravitinos such as
Bose--Fermi supermultiplets of both simple and extended supergravity
in any number of spacetime dimensions, without auxiliary
fields \cite{Nieu81}.

\subsection{From general relativity to supergravity}

It should be stressed that general relativity is naturally related to
supersymmetry, since the requirement of gauge-invariant Rarita--Schwinger
equations \cite{RaSc41} in curved spacetime 
implies Ricci-flatness in four dimensions
\cite{DeZu76}, which is then equivalent to vacuum Einstein equations. 
Of course, despite such a relation does exist, general relativity
can be (and is) formulated without any use of supersymmetry.

The Dirac operator \cite{Espo98} is more fundamental in this framework, since
the $m$-dimensional spacetime metric is entirely re-constructed from the
$\gamma$-matrices, in that 
\begin{equation} 
g^{\mu \nu}=2^{-[m/2]-1} {\rm tr}
(\gamma^{\mu}\gamma^{\nu}+\gamma^{\nu}\gamma^{\mu}).
\label{(16)}
\end{equation}
In four-dimensional spacetime, one can use the tetrad formalism,
with Latin indices $(a,b)$ corresponding to tensors in flat space
(the tangent frames, the freely falling lifts) while Greek indices
$(\mu,\nu)$ correspond to coordinates in curved space. The contravariant 
form of the spacetime metric $g$ is then given by 
\begin{equation}
g^{\mu \nu}=\eta^{ab}e_{a}^{\mu}e_{b}^{\nu},
\end{equation}
where $\eta^{ab}$ is the Minkowski metric and $e_{a}^{\mu}$ are 
the tetrad vectors. The curved-space 
$\gamma$-matrices $\gamma^{\mu}$ are then obtained from the
flat-space $\gamma$-matrices $\gamma^{a}$ and 
from the tetrad according to
\begin{equation}
\gamma^{\mu}=\gamma^{a}e_{a}^{\mu}.
\end{equation}
In Ref. \cite{FrNF76}, the authors assumed that the action functional
describing the interaction of tetrad fields and Rarita--Schwinger
fields in curved spacetime, subject to the Majorana constraint
$\psi_{\rho}(x)=C{\overline \psi}_{\rho}(x)^{T}$, reads as
\begin{equation}
I=\int d^{4}x\biggr[{1\over 4}\kappa^{-2}\sqrt{-g}R
-{1\over 2}\epsilon^{\lambda \rho \mu \nu}
{\overline \psi}_{\lambda}(x)\gamma_{5}\gamma_{\mu}
D_{\nu}\psi_{\rho}(x)\biggr],
\end{equation} 
where the covariant derivative of Rarita--Schwinger fields is
defined by
\begin{equation}
D_{\nu}\psi_{\rho}(x) \equiv \partial_{\nu}\psi_{\rho}(x)
-\Gamma_{\nu \rho}^{\sigma}\psi_{\sigma}
+{1\over 2}\omega_{\nu ab}\sigma^{ab}\psi_{\rho}.
\end{equation}
With a standard notation, $\Gamma_{\nu \rho}^{\sigma}$ are the
Christoffel symbols built from the curved 
spacetime metric $g^{\mu \nu}$, 
$\omega_{\nu ab}$ is the spin-connection (the gauge field associated
to the generators of the Lorentz algebra)
\begin{equation}
\omega_{\nu ab}={1\over 2}\Bigr[e_{a}^{\mu}
(\partial_{\nu}e_{b \mu}-\partial_{\mu}e_{b \nu})
+e_{a}^{\rho}e_{b}^{\sigma}(\partial_{\sigma}e_{c \rho})
e_{\nu}^{c}\Bigr]-(a \leftrightarrow b),
\end{equation}
while $\sigma_{ab}$ is proportional to the commutator of 
$\gamma$-matrices in Minkowski spacetime, i.e.
\begin{equation}
\sigma_{ab} \equiv {1\over 4}[\gamma_{a},\gamma_{b}].
\end{equation}
To investigate the possible supersymmetry possessed by the above
action functional, the authors of Ref. \cite{FrNF76} considered
the transformation laws
\begin{equation}
\delta \psi_{\mu}(x)=\kappa^{-1}D_{\mu}\epsilon(x),
\end{equation}
\begin{equation}
\delta e_{\mu}^{a}(x)={\rm i} \kappa {\overline \epsilon}(x)
\gamma^{a}\psi_{\mu}(x),
\end{equation}
\begin{equation}
\delta g_{\mu \nu}(x)={\rm i}\kappa
{\overline \epsilon}(x) \Bigr[\gamma_{\mu}\psi_{\nu}(x)
+\gamma_{\nu}\psi_{\mu}(x)\Bigr],
\end{equation}
where the supersymmetry parameter is taken to be an arbitrary
Majorana spinor field $\epsilon(x)$ of dimension square root of length.
The assumption of local supersymmetry was non-trivial, and was made
necessary by the coordinate-invariant Lagrangian (i.e. at that stage
one had to avoid, for consistency, the coordinate-dependent 
notion of constant, space-time-independent spinor). After a lengthy
calculation the authors of Ref. \cite{FrNF76} managed to prove full
gauge invariance of the supergravity action. With geometrical hindsight,
one can prove it in a quicker and more elegant way 
by looking at a formulation of
Supergravity as a Yang--Mills Theory \cite{Nieu05}. 

\subsection{New unification}

In modern high energy physics, the emphasis is no longer on fields 
(sections of vector bundles in classical field theory \cite{WaWe90}, 
operator-valued distributions in quantum field theory \cite{Wigh96}), 
but rather on extended objects such as strings
\cite{Deli99}. In string theory, particles are not described as points, but
instead as strings, i.e., one-dimensional extended objects. While a point
particle sweeps out a one-dimensional worldline, the string sweeps out a
worldsheet, i.e., a two-dimensional real surface. For a free string, the
topology of the worldsheet is a cylinder in the case of a closed string, or a
sheet for an open string. It is assumed that different elementary particles
correspond to different vibration modes of the string, in much the same way
as different minimal notes correspond to different vibrational modes of
musical string instruments \cite{Deli99}.
The five different string theories \cite{AnSa02} are
different aspects of a more fundamental unified theory, called
$M$-theory \cite{Beck07}.

In the latest developments, one deals with `branes', which are classical
solutions of the equations of motion of the low-energy string effective
action, that correspond to new non-perturbative states of string theory,
break half of the supersymmetry, and are required by 
duality arguments in theories
with open strings. They have the peculiar property that open strings have
their end-points attached to them \cite{DiLi99a, DiLi99b}.  
Branes have made it possible not only to arrive at the formulation of 
$M$ theory, but also to study perturbative and non-perturbative
properties of the gauge theories living on the world-volume
\cite{DiLi07}. The 
so-called Dirichlet branes \cite{Polc95}, or Dp branes, admit indeed 
two distinct descriptions. On the one hand, they are classical
solutions of the low-energy string effective action 
(as we said before) and may be therefore
described in terms of closed strings. On the other hand, their dynamics
is determined by the degrees of freedom of the open strings with endpoints
attached to their world-volume, satisfying Dirichlet boundary conditions
along the directions transverse to the branes. They may be thus described
in terms of open strings as well. Such a twofold description of Dp branes
laid the foundations of the Maldacena conjecture 
\cite{Mald99} providing the equivalence
between a closed string theory, as the IIB theory on five-dimensional
anti-de Sitter space times the 5-sphere, and $N=4$ super Yang--Mills
with degrees of freedom corresponding to the massless excitations of the
open strings having their endpoints attached to a $D3$ brane. 

For the impact of braneworld picture on phenomenology and unification,
we refer the reader to the seminal work in Refs. \cite{RaSu99a,RaSu99b},
while for the role of extra dimensions in cosmology we should mention
also the work in Refs. \cite{RuSh83a,RuSh83b}. With the language
of pseudo-Riemannian geometry, branes are timelike surfaces embedded into
bulk spacetime \cite{BaNe06,Barv06}. According to this picture, gravity
lives on the bulk, while standard-model gauge fields are confined on the
brane \cite{Barv07}. 
For branes, the normal vector $N$ is spacelike with respect to the
bulk metric $G_{AB}$, i.e.,
\begin{equation}
G_{AB}N^{A}N^{B}=N_{C}N^{C} >0.
\label{(17)}
\end{equation}
For a wide class of brane models, the action functional $S$ 
pertaining to the combined effect of bulk and brane geometry
can be taken to split into the sum \cite{Barv06}
($g_{\alpha \beta}(x)$ being the brane metric)
\begin{equation}
S=S_{4}[g_{\alpha \beta}(x)]+S_{5}[G_{AB}(X)],
\label{(18)}
\end{equation}
while the effective action \cite{DeWi03} $\Gamma$ is formally given by
\begin{equation}
{\rm e}^{{\rm i}\Gamma}=\int DG_{AB}(X) \; 
{\rm e}^{{\rm i}S} \times
{\rm gauge-fixing} \; {\rm term}.
\label{(19)}
\end{equation}
In the functional integral, the gauge-fixed action reads as
(here there is summation as well as integration over repeated indices
\cite{DeWi65, DeWi03, Barv06})
\begin{equation}
S_{{\rm g.f.}}=S_{4}+S_{5}+{1\over 2}F^{A}\Omega_{AB}F^{B}
+{1\over 2}\chi^{\mu}\omega_{\mu \nu}\chi^{\nu},
\label{(20)}
\end{equation}
where $F^{A}$ and $\chi^{\mu}$ are bulk and brane gauge-fixing functionals,
respectively, while $\Omega_{AB}$ and $\omega_{\mu \nu}$ are non-singular
matrices of gauge parameters, similarly to the end of section 4.2.
The gauge-invariance properties of bulk and brane action functionals
can be expressed by saying that there exist vector fields on the
space of histories such that (cf. Eq. (\ref{(11)}))
\begin{equation}
R_{B}S_{5}=0, \; R_{\nu}S_{4}=0,
\label{(21)}
\end{equation}
whose Lie brackets obey a relation formally analogous to 
Eq. (\ref{(12)}) for ordinary type-I theories, i.e.
\begin{equation}
[R_{B},R_{D}]=C_{\; BD}^{A} \; R_{A}, 
\label{(22a)}
\end{equation}
\begin{equation}
[R_{\mu},R_{\nu}]=C_{\; \mu \nu}^{\lambda} \; R_{\lambda}.
\label{(22b)}
\end{equation}
Equations (\ref{(22a)}) and (\ref{(22b)}) refer to the sharply
different Lie algebras of diffeomorphisms on the bulk and the
brane, respectively. The bulk and brane ghost operators 
are therefore
\begin{equation}
Q_{\; B}^{A} \equiv R_{B}F^{A}=F_{\; ,a}^{A} \; R_{\; B}^{a},
\label{(23)}
\end{equation}
\begin{equation}
J_{\; \nu}^{\mu} \equiv R_{\nu}\chi^{\mu}=\chi_{\; ,i}^{\mu} \;
R_{\; \nu}^{i},
\label{(24)}
\end{equation}
respectively, where the commas denote functional differentiation
with respect to the field variables. 
The full bulk integration means integrating first with
respect to all bulk metrics $G_{AB}$ inducing on the boundary
${\partial M}$ the given brane metric $g_{\alpha \beta}(x)$, and then
integrating with respect to all brane metrics.
Thus, one first evaluates the cosmological wave function 
\cite{Barv06} of the bulk
spacetime (which generalizes the wave function of the universe
encountered in canonical quantum gravity), i.e. 
\begin{equation}
\psi_{\rm Bulk}=\int_{G_{AB}[\partial M]=g_{\alpha \beta}}
\mu(G_{AB},S_{C},T^{D}){\rm e}^{{\rm i}{\widetilde S}_{5}},
\label{(25)}
\end{equation}
where $\mu$ is taken to be a suitable measure, the $S_{C},T^{D}$ are
ghost fields, and
\begin{equation}
{\widetilde S}_{5} \equiv S_{5}[G_{AB}]
+{1\over 2}F^{A}\Omega_{AB}F^{B}+S_{A}Q_{\; B}^{A} T^{B}.
\label{(26)}
\end{equation}
Eventually, the effective action results from
\begin{equation}
{\rm e}^{{\rm i}\Gamma}=\int
{\widetilde \mu}(g_{\alpha \beta},\rho_{\gamma},\sigma^{\delta})
{\rm e}^{{\rm i}{\widetilde S}_{4}}\psi_{\rm Bulk},
\label{(27)}
\end{equation}
where ${\widetilde \mu}$ is another putative measure, $\rho_{\gamma}$ and
$\sigma^{\delta}$ are brane ghost fields, and
\begin{equation}
{\widetilde S}_{4} \equiv S_{4}+{1\over 2}\chi^{\mu}\omega_{\mu \nu}
\chi^{\nu}+\rho_{\mu}J_{\; \nu}^{\mu} \sigma^{\nu}.
\label{(28)}
\end{equation}

We would like to stress here that infinite-dimensional manifolds are the
natural arena for studying the quantization of the gravitational field,
even prior to considering a space-of-histories formulation. 
There are, indeed, at least three sources of infinite-dimensionality
in quantum gravity:
\begin{enumerate}
\itemsep=0pt
\item  The infinite-dimensional Lie group (or pseudo-group) of spacetime
diffeomorphisms, which is the invariance group of general relativity
in the first place \cite{DeWi65}, \cite{Stre07}.
\item The infinite-dimensional space of histories in a functional-integral
quantization \cite{DeWi03,DeWi05}.
\item The infinite-dimensional Geroch space of asymptotically simple
spacetimes \cite{Gero71}.
\end{enumerate}

\section{Functional integrals and background fields}

We now study in greater detail some aspects of the use of functional
integrals in quantum gravity, after the previous (formal)
applications to a space of histories formulation.

\subsection{The one-loop approximation}

In the o\-ne-lo\-op ap\-pro\-xi\-ma\-ti\-on (al\-so 
cal\-led sta\-tio\-na\-ry pha\-se
or JWKB me\-th\-od) o\-ne fir\-st ex\-pands {\it both} the met\-ric $g$
and the fields $\phi$ coupled to it about a metric $g_{0}$ and a field
$\phi_{0}$ which are solutions of the classical field equations:
\begin{equation}
g=g_{0}+{\overline g} ,
\label{(3.1.1)}
\end{equation}
\begin{equation}
\phi=\phi_{0}+{\overline \phi} .
\label{(3.1.2)}
\end{equation}
One then assumes that the fluctuations ${\overline g}$ and
${\overline \phi}$ are so small that the dominant 
contribution to the functional integral 
for the in-out amplitude comes from the
quadratic order in the Taylor-series expansion of the action
about the background fields $g_{0}$ and $\phi_{0}$ \cite{Hawk78}:
\begin{equation}
I_{E}[g,\phi]=I_{E}[g_{0},\phi_{0}]+I_{2}[{\overline g},
{\overline \phi}]+ \;
{\rm higher-order} \; {\rm terms} ,
\label{(3.1.3)}
\end{equation}
so that the logarithm of the quantum-gravity amplitude
${\widetilde A}$ can be expressed as 
\begin{equation}
\log \Bigr({\widetilde A}\Bigr) \sim
-I_{E}[g_{0},\phi_{0}]+ \log
\int D[{\overline g},{\overline \phi}]
{\rm e}^{-I_{2}[{\overline g},{\overline \phi}]}.
\label{(3.1.4)}
\end{equation}
It should be stressed that {\it background fields need not be a solution
of any field equation} \cite{DeWi81}, 
but this possibility will not be exploited
in our presentation. For our purposes we are
interested in the second term appearing on the right-hand
side of (\ref{(3.1.4)}). An useful factorization is obtained if
$\phi_{0}$ can be set to zero. One then finds that
$I_{2}[{\overline g},{\overline \phi}]=I_{2}[{\overline g}]
+I_{2}[{\overline \phi}]$, which implies \cite{Hawk79}
\begin{equation}
\log \Bigr({\widetilde A}\Bigr) \sim
-I_{E}[g_{0}]+\log \int D[\phi] {\rm e}^{-I_{2}[\phi]}
+\log \int D[{\overline g}]{\rm e}^{-I_{2}[{\overline g}]}.
\label{(3.1.5)}
\end{equation}
The one-loop term for matter fields with various spins
(and boundary conditions) is extensively studied in
the literature. We here recall some basic results, 
following again Ref. \cite{Hawk79}.

A familiar form of $I_{2}[\phi]$ is 
\begin{equation}
I_{2}[\phi]={1\over 2} \int \phi {\cal B} \phi 
\sqrt{g_{0}} \; d^{4}x ,
\label{(3.1.6)}
\end{equation}
where the elliptic differential operator ${\cal B}$ depends
on the background metric $g_{0}$. Note that ${\cal B}$ is
a second-order operator for bosonic fields, whereas it is
first-order for fermionic fields. In light of (\ref{(3.1.6)}) it is
clear that we are interested in the eigenvalues 
$\Bigr \{ \lambda_{n} \Bigr \}$ of ${\cal B}$, with 
corresponding eigenfunctions $\Bigr \{ \phi_{n} \Bigr \}$. If
boundaries are absent, it is sometimes possible to know
explicitly the eigenvalues with their degeneracies. This is
what happens for example in de Sitter space. If
boundaries are present, however, very little is known about the
detailed form of the eigenvalues, once boundary conditions have
been imposed.

We here assume for simplicity to deal with bosonic fields subject
to (homogeneous) Dirichlet conditions on the boundary surface:
$\phi=0$ on ${\partial M}$, and $\phi_{n}=0$ on ${\partial M}$,
$\forall n$. It is in fact well-known that the Laplace operator
subject to Dirichlet conditions has a positive-definite
spectrum \cite{Chav84}. The field $\phi$ can then be 
expanded in terms of the eigenfunctions $\phi_{n}$ of ${\cal B}$ as 
\begin{equation}
\phi=\sum_{n=n_{0}}^{\infty} y_{n}\phi_{n},
\label{(3.1.7)}
\end{equation}
where the eigenfunctions $\phi_{n}$ are normalized so that 
\begin{equation}
\int \phi_{n} \phi_{m} \sqrt{g_{0}} \; d^{4} x 
=\delta_{nm}.
\label{(3.1.8)}
\end{equation}
Another formula we need is the one expressing the measure
on the space of all fields $\phi$ as 
\begin{equation}
D[\phi]=\prod_{n=n_{0}}^{\infty}\mu \; dy_{n},
\label{(3.1.9)}
\end{equation}
where the normalization parameter $\mu$ has dimensions of
mass or (length)$^{-1}$. Note that, if gauge fields appear 
in the calculation, the choice of gauge-fixing and the form
of the measure in the functional integral are not a trivial problem.

On using well-known results about Gaussian integrals, the 
one-loop amplitudes ${\widetilde A}_{\phi}^{(1)}$
can be now obtained as 
\begin{eqnarray} 
{\widetilde A}_{\phi}^{(1)} & \equiv & \int D[\phi]
{\rm e}^{-I_{2}[\phi]} \nonumber \\
&=& \prod_{n=n_{0}}^{\infty}\int \mu \; dy_{n} \;
{\rm e}^{-{\lambda_{n}\over 2}y_{n}^{2}} \nonumber \\
&=& \prod_{n=n_{0}}^{\infty} 
{\Bigr(2\pi \mu^{2} \lambda_{n}^{-1} \Bigr)}^{1\over 2} \nonumber \\
&=&{1\over \sqrt{{\rm det} \Bigr({1\over 2}\pi^{-1}\mu^{-2}
{\cal B} \Bigr)}}.
\label{(3.1.10)}
\end{eqnarray}

When fermionic fields appear in the functional integral 
for the in-out amplitude,
one deals with a first-order elliptic operator, the Dirac
operator, acting on independent spinor fields $\psi$ and
${\widetilde \psi}$. These are anticommuting Grassmann
variables obeying the Berezin integration rules 
\begin{equation}
\int dw=0 , \int w \; dw=1 .
\label{(3.1.15)}
\end{equation}
The formulae (\ref{(3.1.15)}) are all what we need, since powers
of $w$ greater than or equal to $2$ vanish in light of the
anticommuting property. The reader can then check that the
one-loop amplitude for fermionic fields is 
\begin{equation}
{\widetilde A}_{\psi}^{(1)}={\rm det} 
\left({1\over 2}\mu^{-2}{\cal B}\right).
\label{(3.1.16)}
\end{equation}
The main difference with respect to bosonic fields is the
direct proportionality to the determinant. The following
comments can be useful in understanding the 
meaning of (\ref{(3.1.16)}).

Let us denote again by $\gamma^{\mu}$ the 
curved-space $\gamma$-matrices, and by
$\lambda_{i}$ the eigenvalues of the Dirac operator
$\gamma^{\mu}D_{\mu}$, and suppose that no zero-modes exist.
More precisely, the eigenvalues of $\gamma^{\mu}D_{\mu}$ occur
in equal and opposite pairs: $\pm \lambda_{1}, \pm \lambda_{2}$,
..., whereas the eigenvalues of the Laplace operator on spinors
occur as $(\lambda_{1})^{2}$ twice, $(\lambda_{2})^{2}$ twice,
and so on. For Dirac fermions (D) one thus finds 
\begin{equation}
{\rm det}_{D} \Bigr(\gamma^{\mu}D_{\mu}\Bigr) =
\left(\prod_{i=1}^{\infty}\mid \lambda_{i} \mid \right)
\left(\prod_{i=1}^{\infty}\mid \lambda_{i} \mid \right)
=\prod_{i=1}^{\infty}
{\mid \lambda_{i} \mid}^{2} ,
\label{(3.1.17)}
\end{equation}
whereas in the case of Majorana spinors (M), for which the
number of degrees of freedom is halved, one finds
\begin{equation}
{\rm det}_{M} \Bigr(\gamma^{\mu}D_{\mu}\Bigr)=
\prod_{i=1}^{\infty}\mid \lambda_{i} \mid =
\sqrt{{\rm det}_{D} \Bigr(\gamma^{\mu}D_{\mu} \Bigr)}.
\label{(3.1.18)}
\end{equation}

\subsection{Zeta-function regularization of functional integrals}

The formal expression (\ref{(3.1.10)}) for the one-loop quantum
amplitude clearly diverges since the eigenvalues 
$\lambda_{n}$ increase without bound, and a regularization
is thus necessary. For this purpose, the following technique
has been described and applied by many authors 
\cite{DoCr76,Hawk77,Hawk79}.

Bearing in mind that Riemann's zeta-function $\zeta_{R}(s)$ is
defined as 
\begin{equation}
\zeta_{R}(s) \equiv \sum_{n=1}^{\infty}n^{-s},
\label{(3.2.1)}
\end{equation}
one first defines a generalized (also called spectral) 
zeta-function $\zeta(s)$ 
obtained from the (positive) eigenvalues of the second-order,
self-adjoint operator ${\cal B}$. Such a $\zeta(s)$ can be defined as 
(cf. \cite{Seel67})
\begin{equation}
\zeta(s) \equiv \sum_{n=n_{0}}^{\infty}
\sum_{m=m_{0}}^{\infty} d_{m}(n) \lambda_{n,m}^{-s}.
\label{(3.2.2)}
\end{equation}
This means that all the eigenvalues are completely 
characterized by two integer labels $n$ and $m$, while
their degeneracy $d_{m}$ only depends on $n$. Note that
formal differentiation of (\ref{(3.2.2)}) at the origin yields 
\begin{equation}
{\rm det} \Bigr({\cal B}\Bigr) = {\rm e}^{-\zeta'(0)}.
\label{(3.2.3)}
\end{equation}
This result can be given a sensible meaning since, in four
dimensions, $\zeta(s)$ converges for $Re(s)>2$, and one can
perform its analytic extension to a meromorphic function of
$s$ which only has poles at $s={1\over 2},1,{3\over 2},2$.
Since ${\rm det} \Bigr(\mu {\cal B}\Bigr)=\mu^{\zeta(0)}
{\rm det} \Bigr({\cal B}\Bigr)$, one finds the useful formula 
\begin{equation}
\log \Bigr({\widetilde A}_{\phi}\Bigr)={1\over 2}\zeta'(0)
+{1\over 2} \log \Bigr(2\pi \mu^{2}\Bigr) \zeta(0).
\label{(3.2.4)}
\end{equation}
As we said following (\ref{(3.1.6)}), it may happen quite often that
the eigenvalues appearing in (\ref{(3.2.2)}) are unknown, since the
eigenvalue condition, i.e. the equation leading to the
eigenvalues by virtue of the boundary conditions,
is a complicated equation which cannot
be solved exactly for the eigenvalues. However, since the
scaling properties of the one-loop amplitude are still
given by $\zeta(0)$ (and $\zeta'(0)$) as shown in (\ref{(3.2.4)}),
efforts have been made to compute $\zeta(0)$ 
also in this case. The various steps of this program are as 
follows \cite{Hawk77}.

(1) One first studies the heat equation for the operator
${\cal B}$, i.e.
\begin{equation}
{\partial \over \partial \tau}F(x,y,\tau) + {\cal B}
F(x,y,\tau)=0 ,
\label{(3.2.5)}
\end{equation}
where the Green's function $F$ satisfies the initial
condition $F(x,y,0)=\delta(x,y)$.

(2) Assuming completeness of the set $\Bigr \{ \phi_{n}
\Bigr \}$ of eigenfunctions of ${\cal B}$, the field
$\phi$ can be expanded as
$$
\phi=\sum_{n=n_{i}}^{\infty}a_{n}\phi_{n}.
$$

(3) The Green's function $F(x,y,\tau)$ is then given by 
\begin{equation}
F(x,y,\tau)=\sum_{n=n_{0}}^{\infty}\sum_{m=m_{0}}^{\infty}
{\rm e}^{-\lambda_{n,m}\tau}
\phi_{n,m}(x) \otimes \phi_{n,m}(y).
\label{(3.2.6)}
\end{equation}

(4) The corresponding integrated heat kernel is then 
\begin{equation}
G(\tau) = \int_{M}{\rm Tr} \; F(x,x,\tau)\sqrt{g} \; d^{4}x
=\sum_{n=n_{0}}^{\infty}\sum_{m=m_{0}}^{\infty}
{\rm e}^{-\lambda_{n,m}\tau}.
\label{(3.2.7)}
\end{equation}

(5) In light of (\ref{(3.2.2)}) and (\ref{(3.2.7)}), the generalized 
zeta-function can be also obtained as an integral transform (also
called inverse Mellin transform) of the integrated heat kernel, 
i.e. \cite{Hawk77,Gilk95}
\begin{equation}
\zeta(s)={1\over \Gamma(s)} \int_{0}^{\infty}
\tau^{s-1}G(\tau) \; d\tau.
\label{(3.2.8)}
\end{equation}

(6) The hard part of the analysis is now to prove that 
$G(\tau)$ has an asymptotic expansion as 
$\tau \rightarrow 0^{+}$ \cite{Grei71}. This property has
been proved for all boundary conditions such that the
Laplace operator is self-adjoint and the boundary-value problem is
strongly elliptic \cite{Gilk95,Avra99}. The corresponding 
asymptotic expansion of $G(\tau)$ can be written as 
\begin{equation}
G(\tau) \sim A_{0}\tau^{-2}+A_{1\over 2}\tau^{-{3\over 2}}
+A_{1}\tau^{-1}+A_{3 \over 2}\tau^{-{1\over 2}}+A_{2}+
{\rm O} \Bigr(\sqrt{\tau} \Bigr), 
\label{(3.2.9)}
\end{equation}
which implies 
\begin{equation}
\zeta(0)=A_{2} .
\label{(3.2.10)}
\end{equation}
The result (\ref{(3.2.10)}) is proved by splitting the integral 
in (\ref{(3.2.8)}) into an integral from $0$ to $1$ and an integral from
$1$ to $\infty$. The asymptotic expansion of 
$\int_{0}^{1} \tau^{s-1}G(\tau) \; d\tau$ is then obtained 
by using (\ref{(3.2.9)}).

In other words, for a given second-order self-adjoint elliptic
operator, we study the corresponding heat equation, and the
integrated heat kernel $G(\tau)$. The $\zeta(0)$
value is then given by the constant term appearing in the
asymptotic expansion of $G(\tau)$ as $\tau \rightarrow 0^{+}$. 
The $\zeta(0)$ value also yields the one-loop
divergences of the theory for bosonic and fermionic 
fields \cite{Espo97}.

\subsection{Gravitational instantons}

This section is devoted to the study of the background 
gravitational fields. These gravitational instantons are
complete four-geometries solving the Einstein equations
$R(X,Y)-\Lambda g(X,Y)=0$ when the four-metric $g$ has
signature $+4$ (i.e. it is positive-definite, and thus
called Riemannian). They are of interest because they occur
in the tree-level approximation of the partition function,
and in light of their role in studying tunnelling phenomena.  
Moreover, they can be interpreted as the stationary phase
metrics in the path integrals for the partition functions, $Z$,
of the thermal canonical ensemble and the volume canonical
ensemble. In these cases the action of the instanton gives the
dominant contribution to $- \log Z$. Following \cite{Pope81}, 
essentially three cases can be studied.

\subsubsection{Asymptotically locally Euclidean instantons}

Even though it might seem natural to define first the
asymptotically Euclidean instantons, it turns out that there is
not much choice in this case, since the only asymptotically
Euclidean instanton is flat space. It is in fact well-known
that the action of an asymptotically Euclidean metric with
vanishing scalar curvature is $\geq 0$, and it vanishes if 
and only if the metric is flat. Suppose now that such a metric is a
solution of the Einstein equations $R(X,Y)=0$. Its action
should be thus stationary also under constant conformal
rescalings $g \rightarrow k^{2}g$ of the metric. However,
the whole action rescales then as $I_{E} \rightarrow
k^{2} I_{E}$, so that it can only be stationary and finite
if $I_{E}=0$. By virtue of the theorem previously
mentioned, the metric $g$ must then be flat \cite{GiPo79,LeBr88}.

In the asymptotically locally Euclidean case, however, the
boundary at infinity has topology $S^{3}/ \Gamma$ rather than 
$S^{3}$, where $\Gamma$ is a discrete subgroup of the group $SO(4)$.
Many examples can then be found. The simplest was
discovered by Eguchi and Hanson \cite{EgHa78,EgHa79}, 
and corresponds to
$\Gamma=Z_{2}$ and ${\partial M}=RP^{3}$. This instanton is
conveniently described using three left-invariant 1-forms
$\Bigr \{ \omega_{i} \Bigr \}$ on the 3-sphere, 
satisfying the $SU(2)$ algebra 
$d\omega_{i}=-{1\over 2}\epsilon_{i}^{\; \; jk} \; 
\omega_{j} \wedge \omega_{k}$, and parametrized by Euler
angles as follows:
\begin{equation}
\omega_{1}=(\cos \psi)d\theta + (\sin \psi)(\sin \theta)
d\phi ,
\label{(3.3.1)}
\end{equation}
\begin{equation}
\omega_{2}=-(\sin \psi)d\theta + (\cos \psi)(\sin \theta)
d\phi ,
\label{(3.3.2)}
\end{equation}
\begin{equation}
\omega_{3}=d\psi + (\cos \theta)d\phi, 
\label{(3.3.3)}
\end{equation}
where $\theta \in [0,\pi]$, $\phi \in [0,2\pi]$. The metric of
the Eguchi--Hanson instanton may be thus written in the
Bianchi-IX form \cite{Pope81}
\begin{equation}
g_{1}={\left(1-{a^{4}\over r^{4}}\right)}^{-1}
dr \otimes dr 
+{r^{2}\over 4}
\left[(\omega_{1})^{2}+(\omega_{2})^{2}
+\left(1-{a^{4}\over r^{4}}\right)
(\omega_{3})^{2} \right],
\label{(3.3.4)}
\end{equation}
where $r \in [a,\infty[$. The singularity of $g_{1}$ at
$r=a$ is only a coordinate singularity. We may get rid
of it by defining $4{\rho^{2}\over a^{2}} \equiv
1-{a^{4}\over r^{4}}$, so that, as $r \rightarrow a$, the
metric $g_{1}$ is approximated by the metric 
\begin{equation}
g_{2}=d\rho \otimes d\rho 
+ \rho^{2} {\Bigr [ d\psi + (\cos \theta) d\phi \Bigr]}^{2}
+ {a^{2}\over 4}
\Bigr[ d\theta \otimes d\theta + (\sin \theta)^{2}
d\phi \otimes d\phi \Bigr].
\label{(3.3.5)}
\end{equation}
Regularity of $g_{2}$ at $\rho=0$ is then guaranteed 
provided that one identifies $\psi$ with period $2\pi$.
This implies in turn that the local surfaces $r={\rm constant}$ 
have topology $RP^{3}$ rather than $S^{3}$, as we 
claimed. Note that at $r=a \Rightarrow \rho=0$ the metric
becomes that of a 2-sphere of radius ${a\over 2}$.
Following Ref. \cite{GiHa79}, we say that $r=a$ is
a bolt, where the action of the Killing vector
${\partial \over \partial \psi}$ has a two-dimensional
fixed-point set \cite{Pope81}.

A whole family of multi-instanton solutions is obtained
by taking the group $\Gamma=Z_{k}$. They all have a self-dual
Riemann-curvature tensor, and their metric takes the form 
\begin{equation}
g=V^{-1} {\Bigr(d\tau + {\underline \gamma} \cdot 
{\underline {dx}} \Bigr)}^{2}
+V \; {\underline {dx}} \cdot {\underline {dx}}.
\label{(3.3.6)}
\end{equation}
Following \cite{Pope81}, $V=V({\underline x})$ and
${\underline \gamma}={\underline \gamma}({\underline x})$ on 
an auxiliary flat 3-space with metric
${\underline {dx}} \cdot {\underline {dx}}$. This metric $g$ solves
the Einstein vacuum equations provided that
${\rm grad} \; V = {\rm curl} {\underline \gamma}$, which implies
$\bigtriangleup V=0$. If one takes 
\begin{equation}
V=\sum_{i=1}^{n}{1\over {\mid {\underline x}-
{\underline x_{i}} \mid }},
\label{(3.3.7)}
\end{equation}
one obtains the desired asymptotically locally Euclidean
multi-instantons. In particular, if $n=1$ in (\ref{(3.3.7)}),
$g$ describes flat space, whereas $n=2$ leads to the
Eguchi--Hanson instanton. If $n>2$, there are $(3n-6)$ 
arbitrary parameters, related to the freedom to choose 
the positions ${\underline x_{i}}$ of the singularities
in $V$. These singularities correspond actually to
coordinate singularities in (\ref{(3.3.5)}), and can be removed
by using suitable coordinate transformations \cite{Pope81}.

\subsubsection{Asymptotically flat instantons}

This name is chosen since the underlying idea
is to deal with metrics in the functional integral which
tend to the flat metric in three directions but are
periodic in the Euclidean-time dimension. The basic
example is provided by the Riemannian version 
$g_{R}^{(1)}$ (also called Euclidean) of the
Schwarzschild solution, i.e.
\begin{equation}
g_{R}^{(1)}=\left(1-2{M\over r}\right)
d\tau \otimes d\tau
+{\left(1-2{M\over r}\right)}^{-1}
dr \otimes dr +r^{2} \Omega_{2} ,
\label{(3.3.8)}
\end{equation}
where $\Omega_{2}=d\theta \otimes d\theta + (\sin \theta)^{2}
d\phi \otimes d\phi$ is the metric on a unit 2-sphere.
It is indeed well-known that, in the Lorentzian case,
the metric $g_{L}$ is more conveniently 
written by using Kruskal--Szekeres coordinates 
\begin{equation}
g_{L}=32M^{3}r^{-1}{\rm e}^{-{r\over 2M}}
\Bigr(-dz \otimes dz + dy \otimes dy \Bigr) +r^{2} \Omega_{2} ,
\label{(3.3.9)}
\end{equation}
where $z$ and $y$ obey the relations 
\begin{equation}
-z^{2}+y^{2}= \Bigr({r\over 2M}-1 \Bigr)
{\rm e}^{r\over 2M} ,
\label{(3.3.10)}
\end{equation}
\begin{equation}
{(y+z)\over (y-z)}={\rm e}^{t\over 2M}.
\label{(3.3.11)}
\end{equation}
In the Lorentzian case, the coordinate singularity
at $r=2M$ can be thus avoided, whereas the curvature
singularity at $r=0$ remains and is described by the
surface $z^{2}-y^{2}=1$. However, if we set 
$\zeta={\rm i}z$, the analytic continuation to the section
of the complexified space-time where $\zeta$ is real
yields the positive-definite (i.e. Riemannian) metric 
\begin{equation}
g_{R}^{(2)} = 32M^{3}r^{-1}{\rm e}^{-{r\over 2M}}
\Bigr(d\zeta \otimes d\zeta + dy \otimes dy \Bigr) 
+r^{2} \Omega_{2} ,
\label{(3.3.12)}
\end{equation}
where 
\begin{equation}
\zeta^{2}+y^{2}= \Bigr({r\over 2M}-1 \Bigr) {\rm e}^{r\over 2M}.
\label{(3.3.13)}
\end{equation}
It is now clear that also the curvature singularity
at $r=0$ has disappeared, since the left-hand side of
(\ref{(3.3.13)}) is $\geq 0$, whereas the right-hand side of
(\ref{(3.3.13)}) would be equal to $-1$ at $r=0$. Note also
that, by setting $z=-{\rm i} \zeta$ and 
$t=-{\rm i} \tau$ in (\ref{(3.3.11)}),
and writing $\zeta^{2}+y^{2}$ as 
$(y+{\rm i}\zeta)(y-{\rm i}\zeta)$ in
(\ref{(3.3.13)}), one finds 
\begin{equation}
y+{\rm i}\zeta={\rm e}^{{\rm i}\tau \over 4M}
\sqrt{{r\over 2M}-1} \; \; {\rm e}^{r\over 4M},
\label{(3.3.14)}
\end{equation}
\begin{equation}
y=\cos \Bigr({\tau \over 4M}\Bigr)
\sqrt{{r\over 2M}-1} \; \; {\rm e}^{r\over 4M},
\label{(3.3.15)}
\end{equation}
which imply that the Euclidean time $\tau$ is periodic
with period $8\pi M$. This periodicity on the Euclidean
section leads to the interpretation of the Riemannian
Schwarzschild solution as describing a black hole in
thermal equilibrium with gravitons at a temperature
$(8 \pi M)^{-1}$ \cite{Pope81}. Moreover, the fact that
any matter-field Green's function on this Schwarzschild
background is also periodic in imaginary time leads to 
some of the thermal-emission properties of black holes.
This is one of the greatest conceptual revolutions in
modern gravitational physics.

Interestingly, a new asymptotically flat gravitational instanton
has been found by Chen and Teo in Ref. \cite{ChTe11}. It has
an $U(1) \times U(1)$ isometry group and some novel global 
features with respect to the other two asymptotically flat
instantons, i.e. Euclidean Schwarzschild and Euclidean Kerr.

There is also a local version of the asymptotically flat
boundary condition in which ${\partial M}$ has the
topology of a non-trivial $S^{1}$-bundle over $S^{2}$,
i.e. $S^{3} / \Gamma$, where $\Gamma$ is a discrete
subgroup of $SO(4)$. However, unlike the asymptotically
Euclidean boundary condition, the $S^{3}$ is distorted
and expands with increasing radius in only two directions
rather than three \cite{Pope81}. The simplest example of an
asymptotically locally flat instanton is the self-dual
(i.e. with self-dual curvature 2-form)
Taub-NUT solution, which can be regarded as a special
case of the two-parameter Taub-NUT metrics 
\begin{equation}
g={(r+M)\over (r-M)} dr \otimes dr +
4M^{2}{(r-M)\over (r+M)}
(\omega_{3})^{2} 
+\Bigr(r^{2}-M^{2}\Bigr)
\Bigr[(\omega_{1})^{2}+(\omega_{2})^{2}\Bigr],
\label{(3.3.16)}
\end{equation}
where the $\Bigr \{ \omega_{i} \Bigr \}$ have been defined in 
(\ref{(3.3.1)})--(\ref{(3.3.3)}). The main properties of the metric
(\ref{(3.3.16)}) are 

(I) $r \in [M,\infty[$, and $r=M$ is a removable
coordinate singularity provided that $\psi$ 
is identified modulo $4\pi$;

(II) the $r={\rm constant}$ surfaces have $S^{3}$ topology;

(III) $r=M$ is a point at which the isometry generated
by the Killing vector ${\partial \over \partial \psi}$
has a zero-dimensional fixed-point set.
\vskip 0.3cm
\noindent
In other words, $r=M$ is a nut, using the terminology
in Ref. \cite{GiHa79}. 

There is also a family of asymptotically locally flat 
multi-Taub-NUT instantons. Their metric takes the form
(\ref{(3.3.6)}), but one should bear in mind that the formula
(\ref{(3.3.7)}) is replaced by 
\begin{equation}
V=1+\sum_{i=1}^{n} {2M\over {\mid {\underline x}
-{\underline x_{i}} \mid}}.
\label{(3.3.17)}
\end{equation}
Again, the singularities at ${\underline x}=
{\underline x_{i}}$ can be removed, and the 
instantons are all self-dual.

\subsubsection{Compact instantons}

Compact gravitational instantons occur in the course
of studying the topological structure of the gravitational
vacuum. This can be done by first of all normalizing all
metrics in the functional integral to have a given
4-volume $V$, and then evaluating the instanton
contributions to the partition function as a function
of their topological complexity. One then sends the volume
$V$ to infinity at the end of the calculation. If one wants
to constrain the metrics in the functional integral to have a
volume $V$, this can be obtained by adding a term
${\Lambda \over 8\pi}V$ to the action. The stationary points
of the modified action are solutions of the Einstein equations
with cosmological constant $\Lambda$, i.e.
$R(X,Y)-\Lambda g(X,Y)=0$.
\vskip 0.3cm
\noindent
The few compact instantons that are known can be described
as follows \cite{Pope81}.

(1) The 4-sphere $S^{4}$, i.e. the Riemannian version of
de Sitter space obtained by analytic continuation to
positive-definite metrics. Setting to $3$ for convenience
the cosmological constant, the metric on $S^{4}$ takes the
form \cite{Pope81} 
\begin{equation}
g_{I}=d\beta \otimes d\beta +{1\over 4}(\sin \beta)^{2}
\Bigr[(\omega_{1})^{2}+(\omega_{2})^{2}+
(\omega_{3})^{2} \Bigr],
\label{(3.3.18)}
\end{equation}
where $\beta \in [0,\pi]$. The apparent singularities at
$\beta=0,\pi$ can be made into regular nuts, provided that
the Euler angle $\psi$ is identified modulo $4\pi$. The 
$\beta={\rm constant}$ surfaces are topologically 
$S^{3}$, and the isometry
group of the metric (\ref{(3.3.18)}) is $SO(5)$.

(2) If in $C^{3}$ we identify $(z_{1},z_{2},z_{3})$ and
$(\lambda z_{1}, \lambda z_{2}, \lambda z_{3})$,
$\forall \lambda \in C - \{ 0 \}$, we obtain, by 
definition, the complex projective space $CP^{2}$. 
For this two-dimensional complex
space one can find a real four-dimensional metric, which
solves the Einstein equations with cosmological constant
$\Lambda$. If we set $\Lambda$ to $6$ for convenience, the
metric of $CP^{2}$ takes the form \cite{Pope81} 
\begin{equation}
g_{II}=d\beta \otimes d\beta +{1\over 4}(\sin \beta)^{2}
\Bigr[(\omega_{1})^{2}+(\omega_{2})^{2}
+(\cos \beta)^{2}(\omega_{3})^{2} \Bigr],
\label{(3.3.19)}
\end{equation}
where $\beta \in \Bigr[0,{\pi \over 2} \Bigr]$. A bolt
exists at $\beta={\pi\over 2}$, where
${\partial \over \partial \psi}$ has a two-dimensional
fixed-point set. The isometry group of $g_{II}$ is locally
$SU(3)$, which has a $U(2)$ subgroup acting on the three-spheres
$\beta={\rm constant}$.

(3) The Einstein metric on the product manifold
$S^{2} \times S^{2}$ is obtained as the direct sum of the
metrics on two 2-spheres, i.e.
\begin{equation}
g={1\over \Lambda}\sum_{i=1}^{2}
\Bigr(d\theta_{i} \otimes d\theta_{i} +
(\sin \theta_{i})^{2} d\phi_{i} \otimes d\phi_{i}\Bigr) .
\label{(3.3.20)}
\end{equation}
The metric (\ref{(3.3.20)}) is invariant under the 
$SO(3) \times SO(3)$ isometry group of
$S^{2} \times S^{2}$, but is not of Bianchi-IX type
as (\ref{(3.3.18)})-(\ref{(3.3.19)}). 
This can be achieved by a coordinate
transformation leading to \cite{Pope81} 
\begin{equation}
g_{III}= d\beta \otimes d\beta +(\cos \beta)^{2}
(\omega_{1})^{2}+(\sin \beta)^{2}(\omega_{2})^{2}+(\omega_{3})^{2} ,
\label{(3.3.21)}
\end{equation}
where $\Lambda=2$ and $\beta \in \Bigr [0,{\pi \over 2}
\Bigr]$. Regularity at $\beta=0,{\pi \over 2}$ is obtained
provided that $\psi$ is identified modulo $2\pi$
(cf. (\ref{(3.3.18)})). Remarkably, this is a regular Bianchi-IX
Einstein metric in which the coefficients of
$\omega_{1},\omega_{2}$ and $\omega_{3}$ are all different.

(4) The nontrivial $S^{2}$-bundle over $S^{2}$ has a metric
which, by setting $\Lambda=3$, may be cast in the form 
\cite{Page78a,Pope81} 
\begin{equation}
g_{IV}=(1+\nu^{2})\Bigr[f_{1}(x)dx \otimes dx
+f_{2}(x)\Bigr((\omega_{1})^{2}+(\omega_{2})^{2}\Bigr)
+f_{3}(x)(\omega_{3})^{2}\Bigr],
\label{(78)}
\end{equation}
 where $x \in [0,1]$, $\nu$ is the positive root of 
\begin{equation}
w^{4}+4w^{3}-6w^{2}+12w-3=0,
\label{(3.3.23)}
\end{equation}
and the functions $f_{1},f_{2},f_{3}$ are defined by
\begin{equation}
f_{1}(x) \equiv {(1-\nu^{2}x^{2})\over 
(3-\nu^{2}-\nu^{2}(1+\nu^{2})x^{2})(1-x^{2})},
\label{(79)}
\end{equation}
\begin{equation}
f_{2}(x) \equiv {(1-\nu^{2}x^{2})\over
(3+6\nu^{2}-\nu^{4})},
\label{(80)}
\end{equation}
\begin{equation}
f_{3}(x) \equiv {(3-\nu^{2}-\nu^{2}(1+\nu^{2})x^{2})(1-x^{2})
\over (3-\nu^{2})(1-\nu^{2}x^{2})}.
\label{(81)}
\end{equation}
The isometry group corresponding to (\ref{(78)}) may be
shown to be $U(2)$.

(5) Another compact instanton of fundamental importance
is the $K3$ surface, whose explicit metric has not yet
been found. $K3$ is defined as the compact complex surface
whose first Betti number and first Chern class are vanishing.
A physical picture of the K3 gravitational instanton has been
obtained by Page \cite{Page78b}.
\vskip 0.3cm
\noindent
Two topological invariants exist which may be used to
characterize the various gravitational instantons studied
so far. These invariants are the Euler number $\chi$ and the
Hirzebruch signature $\tau$. The Euler number can be defined
as an alternating sum of Betti numbers, i.e.
\begin{equation}
\chi \equiv  B_{0}-B_{1}+B_{2}-B_{3}+B_{4}. 
\label{(3.3.24)}
\end{equation}
The Hirzebruch signature can be defined as 
\begin{equation}
\tau \equiv B_{2}^{+}-B_{2}^{-}, 
\label{(3.3.25)}
\end{equation}
where $B_{2}^{+}$ is the number of self-dual harmonic
2-forms, and $B_{2}^{-}$ is the number of
anti-self-dual harmonic 2-forms [in terms of the Hodge-star
operator ${ }^{*}F_{ab} \equiv {1\over 2}\epsilon_{abcd}F^{cd}$,
self-duality of a 2-form $F$ is expressed as
${ }^{*}F=F$, and anti-self-duality as ${ }^{*}F=-F$].
In the case of compact four-dimensional manifolds
without boundary, $\chi$ and $\tau$
can be expressed as integrals of the curvature \cite{Hawk79}
\begin{equation}
\chi={1\over 128 \pi^{2}}
\int_{M} R_{\lambda \mu \nu \rho} \; R_{\alpha \beta \gamma \delta} \; 
\epsilon^{\lambda \mu \alpha \beta } \; \epsilon^{\nu \rho \gamma \delta}
\; \sqrt{g} \; d^{4}x, 
\label{(3.3.26)}
\end{equation}
\begin{equation}
\tau={1\over 96 \pi^{2}}
\int_{M} R_{\lambda \mu \nu \rho} \; 
R_{\; \; \; \alpha \beta}^{\lambda \mu} \;
\epsilon^{\nu \rho \alpha \beta} \; \sqrt{g} \; d^{4}x.  
\label{(3.3.27)}
\end{equation}
For the instantons previously listed one finds \cite{Pope81} 
\vskip 0.3cm
\noindent
Eguchi--Hanson: $\chi=2$, $\tau=1$.
\vskip 0.3cm
\noindent
Asymptotically locally Euclidean multi-instantons:
$\chi=n$, $\tau=n-1$.
\vskip 0.3cm
\noindent
Schwarzschild: $\chi=2$, $\tau=0$.
\vskip 0.3cm
\noindent
Taub-NUT: $\chi=1$, $\tau=0$.
\vskip 0.3cm
\noindent
Asymptotically locally flat multi-Taub-NUT instantons:
$\chi=n$, $\tau=n-1$.
\vskip 0.3cm
\noindent
$S^{4}$: $\chi=2$, $\tau=0$.
\vskip 0.3cm
\noindent
$CP^{2}$: $\chi=3$, $\tau=1$.
\vskip 0.3cm
\noindent
$S^{2} \times S^{2}$: $\chi=4$, $\tau=0$.
\vskip 0.3cm
\noindent
$S^{2}$-bundle over $S^{2}$: $\chi=4$, $\tau=0$.
\vskip 0.3cm
\noindent
$K3$: $\chi=24$, $\tau=16$.

\section{Spectral zeta-functions in one-loop quantum cosmology}

In the late nineties a systematic investigation of boundary conditions
in quantum field theory and quantum gravity has been performed
(see Refs. \cite{Luck91,Vass95,Espo97,MoSi97,Avra99} and the many 
references therein). It is now
clear that the set of fully gauge-invariant boundary conditions
in quantum field theory, providing a unified scheme for Maxwell,
Yang--Mills and General Relativity is as follows:
\begin{equation}
\Bigr[\pi {\cal A}\Bigr]_{\partial M}=0,
\label{(84)}
\end{equation}
\begin{equation}
\Bigr[\Phi({\cal A})\Bigr]_{\partial M}=0,
\label{(85)}
\end{equation}
\begin{equation}
[\varphi]_{\partial M}=0,
\label{(86)}
\end{equation}
where $\pi$ is a projection operator, ${\cal A}$ is the Maxwell
potential, or the Yang--Mills potential, or the metric (more
precisely, their perturbation about a background value which can be
set to zero for Maxwell or Yang--Mills), $\Phi$ is the gauge-fixing
functional, $\varphi$ denotes the set of ghost fields for these
bosonic theories. Equations (\ref{(84)}) and (\ref{(85)}) are both
preserved under infinitesimal gauge transformations provided that the
ghost obeys homogeneous Dirichlet conditions as in Eq. (\ref{(86)}).
For gravity, it may be convenient to choose $\Phi$ so as to have an
operator $P$ of Laplace type in the Euclidean theory.

\subsection{Eigenvalue condition for scalar modes}

In a quantum theory of the early universe via functional integrals,
the semiclassical analysis remains a valuable tool, but the tree-level
approximation might be an oversimplification. Thus, it seems
appropriate to consider at least the one-loop approximation.
On the portion of flat Euclidean 4-space bounded by a 3-sphere,
called Euclidean 4-ball and relevant for one-loop quantum
cosmology \cite{HaHa83,Hawk84,Espo97} when a portion of 4-sphere
bounded by a 3-sphere is studied in the limit of small  
3-geometry \cite{Schl85}, the metric perturbations
$h_{\mu \nu}$ can be expanded in terms of scalar, transverse vector,
transverse-traceless tensor harmonics on 
the 3-sphere $S^{3}$ of radius $a$. For
vector, tensor and ghost modes, boundary conditions reduce to 
Dirichlet or Robin. For scalar modes, one finds eventually the
eigenvalues $E=X^{2}$ from the roots $X$ of \cite{Espo05a,Espo05b} 
\begin{equation}
J_{n}'(x) \pm {n \over x}J_{n}(x)=0,
\label{(87)}
\end{equation}
\begin{equation}
J_{n}'(x)+\left(-{x \over 2} \pm {n \over x}\right)J_{n}(x)=0,
\label{(88)}
\end{equation}
where $J_{n}$ are the Bessel functions of first kind.
Note that both $x$ and $-x$ solve the same equation.

\subsection{Four spectral zeta-functions for scalar modes}

By virtue of the Cauchy theorem and of suitable rotations of
integration contours in the complex plane \cite{BGKE96},
the eigenvalue conditions (\ref{(87)}) and (\ref{(88)}) give
rise to the following four spectral zeta-functions 
\cite{Espo05a,Espo05b}:
\begin{equation}
\zeta_{A,B}^{\pm}(s) \equiv {\sin(\pi s) \over \pi} 
\sum_{n=3}^{\infty}n^{-(2s-2)}
\int_{0}^{\infty}dz {{\partial \over \partial z}
\log F_{A,B}^{\pm}(zn) \over z^{2s}},
\label{(89)}
\end{equation}
where, denoting by $I_{n}$ the modified Bessel functions of 
first kind (here $\beta_{+} \equiv n, \beta_{-} \equiv n+2$),
\begin{equation}
F_{A}^{\pm}(zn) \equiv z^{-\beta_{\pm}}
\Bigr(zn I_{n}'(zn) \pm n I_{n}(zn)\Bigr),
\label{(90)}
\end{equation}
\begin{equation}
F_{B}^{\pm}(zn) \equiv z^{-\beta_{\pm}}\left(zn I_{n}'(zn)
+\left({z^{2}n^{2}\over 2} \pm n \right)I_{n}(zn)\right).
\label{(91)}
\end{equation}
Regularity at the origin is easily proved in the elliptic sectors,
corresponding to $\zeta_{A}^{\pm}(s)$ and $\zeta_{B}^{-}(s)$
\cite{Espo05a,Espo05b}.

\subsection{Regularity at the origin of $\zeta_{B}^{+}$}

With the notation in Refs. \cite{Espo05a,Espo05b}, if one defines
the variable $\tau \equiv (1+z^{2})^{-{1\over 2}}$, one can write
the uniform asymptotic expansion of $F_{B}^{+}$ in the form
\cite{Espo05a,Espo05b}
\begin{equation}
F_{B}^{+} \sim {{\rm e}^{n \eta(\tau)}\over h(n)\sqrt{\tau}}
{(1-\tau^{2})\over \tau} \left(1+\sum_{j=1}^{\infty}
{r_{j,+}(\tau)\over n^{j}}\right). 
\label{(92)}
\end{equation}
On splitting the integral $\int_{0}^{1}d\tau=\int_{0}^{\mu}d\tau
+\int_{\mu}^{1}d\tau$ with $\mu$ small, one gets an asymptotic expansion 
of the left-hand side of Eq. (\ref{(89)}) by writing, in the first
interval on the right-hand side, 
\begin{equation}
\log \left(1+\sum_{j=1}^{\infty}{r_{j,+}(\tau)\over n^{j}}\right)
\sim \sum_{j=1}^{\infty}{R_{j,+}(\tau)\over n^{j}},
\label{(93)}
\end{equation}
and then computing \cite{Espo05a,Espo05b}
\begin{equation}
C_{j}(\tau) \equiv {\partial R_{j,+}\over \partial \tau}
=(1-\tau)^{-j-1}\sum_{a=j-1}^{4j}K_{a}^{(j)}\tau^{a}.
\label{(94)}
\end{equation}
Remarkably, by virtue of the identity obeyed by the spectral
coefficients $K_{a}^{(j)}$ on the 4-ball, i.e. 
\begin{equation}
g(j) \equiv \sum_{a=j}^{4j}{\Gamma(a+1)\over \Gamma(a-j+1)}
K_{a}^{(j)}=0,
\label{(95)}
\end{equation}
which holds $\forall j=1,...,\infty$, one finds \cite{Espo05a,Espo05b}
\begin{equation}
\lim_{s \to 0} s \zeta_{B}^{+}(s)={1\over 6}\sum_{a=3}^{12}
a(a-1)(a-2)K_{a}^{(3)}=0,
\label{(96)}
\end{equation}
and \cite{Espo05a,Espo05b}
\begin{equation}
\zeta_{B}^{+}(0)={5\over 4}+{1079 \over 240}-{1\over 2}
\sum_{a=2}^{12}\omega(a)K_{a}^{(3)}+\sum_{j=1}^{\infty}f(j)g(j)
={296\over 45},
\label{(97)}
\end{equation}
where, on denoting here by $\psi$ the logarithmic
derivative of the $\Gamma$-function \cite{Espo05a,Espo05b},
\begin{eqnarray}
\; & \; &
\omega(a) \equiv {1\over 6}{\Gamma(a+1)\over \Gamma(a-2)}
\biggr[-\log(2)-{(6a^{2}-9a+1)\over 4}{\Gamma(a-2)\over \Gamma(a+1)}
\nonumber \\
&+& 2\psi(a+1)-\psi(a-2)-\psi(4)\biggr],
\label{(98)}
\end{eqnarray}
\begin{equation}
f(j) \equiv {(-1)^{j}\over j!}\Bigr[-1-2^{2-j}
+\zeta_{R}(j-2)(1-\delta_{j,3})+\gamma \delta_{j,3}\Bigr].
\label{(99)}
\end{equation}
Equation (\ref{(95)}) achieves three goals:
\vskip 0.3cm
\noindent
(i) Vanishing of $\log(2)$ coefficient in (\ref{(97)});
\vskip 0.3cm
\noindent
(ii) Vanishing of $\sum_{j=1}^{\infty}f(j)g(j)$ in (\ref{(97)});
\vskip 0.3cm
\noindent
(iii) Regularity at the origin of $\zeta_{B}^{+}$.

\subsection{Interpretation of the result}

Since all other $\zeta(0)$ values for pure gravity obtained in the
literature are negative,
the analysis here briefly outlined shows that only fully
diffeomorphism-invariant boundary conditions lead to a positive
$\zeta(0)$ value for pure gravity on the 4-ball, and hence {\it only
fully diffeomorphism-invariant boundary conditions lead to a
vanishing cosmological wave function for vanishing 3-geometries
at one-loop level, at least on the Euclidean 4-ball}. If the 
probabilistic interpretation is tenable for the whole universe,
this means that {\it the universe has vanishing probability of reaching
the initial singularity} at $a=0$, 
which is therefore avoided by virtue
of quantum effects \cite{Espo05a,Espo05b}, since the one-loop
wave function is proportional to $a^{\zeta(0)}$ \cite{Schl85}.

Interestingly, quantum cosmology can have observational consequences
as well. For example, the work in Ref. \cite{KiKr11} has derived the
primordial power spectrum of density fluctuations in the framework of
quantum cosmology, by performing a Born--Oppenheimer approximation
of the Wheeler--DeWitt equation for an inflationary universe with a
scalar field. In this way one first recovers the scale-invariant power
spectrum that is found as an approximation in the simplest inflationary
models. One then obtains quantum gravitational corrections to this
spectrum, discussing whether they lead to measurable signatures in the
Cosmic Microwave Background anisotropy spectrum \cite{KiKr11}.

\section{Hawking's radiation}

Hawking's theoretical discovery of particle creation by black holes
\cite{Hawk74, Hawk75} has led, 
along the years, to many important developments in quantum
field theory in curved spacetime, quantum gravity and string theory.
Thus, we devote this section to a brief review of such an effect,
relying upon the DAMTP lecture notes by Townsend \cite{Town97}.
For this purpose, we consider a massless scalar field 
$\Phi$ in a Schwarzschild black hole spacetime. The positive-frequency
outgoing modes of $\Phi$ are known to behave, near future null
infinity ${\cal F}^{+}$, as
\begin{equation}
\Phi_{\omega} \sim {\rm e}^{-{\rm i}\omega u}.
\label{(500)}
\end{equation}
According to a geometric optics approximation, a particle's worldline
is a null ray $\gamma$ of constant phase $u$, and we trace this ray
backwards in time from ${\cal F}^{+}$. The later it reaches
${\cal F}^{+}$, the closer it must approach the future event horizon
${\cal H}^{+}$ in the exterior spacetime before entering the star.  
The ray $\gamma$ belongs to a family of rays whose limit as 
$t \rightarrow \infty$ is a null geodesic generator, denoted by
$\gamma_{H}$, of ${\cal H}^{+}$. One can specify $\gamma$ by giving
its affine distance from $\gamma_{H}$ along an ingoing null geodesic
passing through ${\cal H}^{+}$. The affine parameter on this ingoing
null geodesic is $U$, so $U=-\epsilon$. One can thus write,
on $\gamma$ near ${\cal H}^{+}$ ($\kappa$ being the surface gravity),
\begin{equation}
u=-{1\over \kappa}\log \epsilon,
\label{(501)}
\end{equation}
so that positive-frequency outgoing modes have, near ${\cal H}^{+}$,
the approximate form
\begin{equation}
\Phi_{\omega} \sim {\rm e}^{{{\rm i}\omega \over \kappa}\log
\epsilon}.
\label{(502)}
\end{equation}
This describes increasingly rapid oscillations as 
$\epsilon \rightarrow 0$, and hence the geometric optics approximation
is indeed justified at late times.

The positive-frequency outgoing modes should be matched onto a
solution of the Klein--Gordon equation near past null infinity
${\cal F}^{-}$. When geometric optics holds, one performs parallel
transport of the vectors $n$ parallel to ${\cal F}^{-}$ and $l$
orthogonal to $n$ back to ${\cal F}^{-}$ along
the continuation of $\gamma_{H}$. Such a continuation can be taken
to meet ${\cal F}^{-}$ at $v=0$. The continuation of the null ray
$\gamma$ back to ${\cal F}^{-}$ meets ${\cal F}^{-}$ at an affine
distance $\epsilon$ along an outgoing null geodesic on ${\cal F}^{-}$.
The affine parameter on outgoing null geodesics in ${\cal F}^{-}$
is $v$, because the line element takes on ${\cal F}^{-}$ the form
\begin{equation}
ds^{2}=du \; dv +r^{2}d\Omega^{2},
\label{(503)}
\end{equation}
$d\Omega^{2}$ being the line element on a unit 2-sphere, so that
$v=-\epsilon$ and
\begin{equation}
\Phi_{\omega} \sim {\rm e}^{{{\rm i}\omega \over \kappa}
\log(-v)}.
\label{(504)}
\end{equation}
This holds for negative values of $v$. When $v$ is instead positive,
an ingoing null ray from ${\cal F}^{-}$ passes through ${\cal H}^{+}$
and does not reach ${\cal F}^{+}$, hence the positive-frequency 
outgoing modes depend on $v$ on ${\cal F}^{-}$, where
\begin{equation}
\Phi_{\omega}(v)=0 \; {\rm if} \; v>0, \;
{\rm e}^{{{\rm i}\omega \over \kappa}\log(-v)} \;
{\rm if} \; v<0.
\label{(505)}
\end{equation}
Consider now the Fourier transform
\begin{eqnarray}
{\widetilde \Phi}_{\omega}& \equiv & \int_{-\infty}^{\infty}
{\rm e}^{{\rm i}\omega' v}\Phi_{\omega}(v)dv \nonumber \\
&=& \int_{-\infty}^{0}
{\rm e}^{{\rm i}\omega' v +{{\rm i}\omega \over \kappa}
\log(-v)} dv.
\label{(506)}
\end{eqnarray}
In this integral, let us choose the branch cut in the complex
$v$-plane to lie along the real axis. For positive $\omega'$
let us rotate contour to the positive imaginary axis and then
set $v={\rm i}x$ to get
\begin{eqnarray}
{\widetilde \Phi}_{\omega}(\omega')&=& -{\rm i}
\int_{0}^{\infty}
{\rm e}^{-\omega' x+{{\rm i}\omega \over \kappa}
\log \Bigr(x {\rm e}^{-{\rm i}\pi /2}\Bigr)}dx \nonumber \\
&=&-{\rm e}^{\pi \omega \over 2 \kappa}
\int_{0}^{\infty}
{\rm e}^{-\omega' x +{{\rm i}\omega \over \kappa}
\log(x)}dx.
\label{(507)}
\end{eqnarray}
Since $\omega'$ is positive the integral converges. When $\omega'$
is negative one can rotate the contour to the negative imaginary 
axis and then set $v=-{\rm i}x$ to get
 \begin{eqnarray}
{\widetilde \Phi}_{\omega}(\omega')&=& {\rm i}
\int_{0}^{\infty}
{\rm e}^{\omega' x+{{\rm i}\omega \over \kappa}
\log \Bigr(x {\rm e}^{{\rm i}\pi /2}\Bigr)}dx \nonumber \\
&=&{\rm e}^{-{\pi \omega \over 2 \kappa}}
\int_{0}^{\infty}
{\rm e}^{\omega' x +{{\rm i}\omega \over \kappa}
\log(x)}dx.
\label{(508)}
\end{eqnarray}
From the two previous formulae one gets
\begin{equation}
{\widetilde \Phi}_{\omega}(-\omega')
=-{\rm e}^{-{\pi \omega \over \kappa}}
{\widetilde \Phi}_{\omega}(\omega') \; {\rm if} \; \omega' >0.
\label{(509)}
\end{equation}
Thus, a mode of positive frequency $\omega$ on ${\cal F}^{+}$
matches, at late times, onto positive and negative modes on
${\cal F}^{-}$. For positive $\omega'$ one can identify
\begin{equation}
A_{\omega \omega'}={\widetilde \Phi}_{\omega}(\omega'),
\label{(510)}
\end{equation}
\begin{equation}
B_{\omega \omega'}={\widetilde \Phi}_{\omega}(-\omega')
=-{\rm e}^{-{\pi \omega \over \kappa}}
{\widetilde \Phi}_{\omega}(\omega'),
\label{(511)}
\end{equation}
as the Bogoliubov coefficients. These formulae imply that
\begin{equation}
B_{ij}=-{\rm e}^{-{\pi \omega\over \kappa}}A_{ij}.
\label{(512)}
\end{equation}
On the other hand, the matrices $A$ and $B$ should satisfy the
Bogoliubov relations, from which
\begin{eqnarray}
\delta_{ij}&=& (AA^{\dagger}-BB^{\dagger})_{ij} \nonumber \\
&=& \sum_{l}(A_{il}A_{jl}^{*}-B_{il}B_{jl}^{*}) \nonumber \\
&=& \Bigr[{\rm e}^{{\pi (\omega_{i}+\omega_{j})\over \kappa}}-1 \Bigr]
\sum_{l}B_{il}B_{jl}^{*},
\label{(513)}
\end{eqnarray}
where we have inserted the formula relating $B_{ij}$ to $A_{ij}$.
Now one can take $i=j$ to get
\begin{equation}
(BB^{\dagger})_{ii}={1\over {\rm e}^{{2\pi \omega_{i} \over \kappa}}
-1}.
\label{(514)}
\end{equation}
Eventually, one needs the inverse Bogoliubov coefficients corresponding
to a positive-frequency mode on ${\cal F}^{-}$ matching onto positive-
and negative-frequency modes on ${\cal F}^{+}$. Since the inverse $B$
coefficient is found to be
\begin{equation}
B'=-B^{T},
\label{(515)}
\end{equation}
the late-time particle flux through ${\cal F}^{+}$, given a vacuum
on ${\cal F}^{-}$, turns out to be
\begin{equation}
\langle N_{i} \rangle_{{\cal F}^{+}}=\Bigr((B')^{\dagger}B'\Bigr)_{ii}
=\Bigr(B^{*}B^{T}\Bigr)_{ii}=\Bigr(BB^{T}\Bigr)_{ii}^{*}.
\label{(516)}
\end{equation}
From reality of $(BB^{T})_{ii}$, the previous formulae lead to
\begin{equation}
\langle N_{i} \rangle_{{\cal F}^{+}}
={1\over {\rm e}^{{2 \pi \omega_{i}\over \kappa}}-1}.
\label{(517)}
\end{equation}
Remarkably, this is the Planck distribution for black body radiation
from a Schwarzschild black hole at the Hawking temperature
\begin{equation}
T_{H}={\hbar \kappa \over 2\pi}.
\label{(518)}
\end{equation}

\section{Achievements and open problems}

At this stage, the general reader might well be wondering what has
been gained by working on the quantum gravity problem over so many
decades. Indeed, at the theoretical level, at least the following
achievements can be brought to his (or her) attention:
\vskip 0.3cm
\noindent
(i) The ghost fields \cite{Feyn63} 
necessary for the functional-integral quantization
of gravity and Yang--Mills theories \cite{DeWi67b,FaPo67} 
have been discovered, jointly with
a deep perspective on the space of histories formulation.
\vskip 0.3cm
\noindent
(ii) The Vilkovisky--DeWitt gauge-invariant effective action 
\cite{Vilk84,DeWi03}
has been obtained and thoroughly studied.
\vskip 0.3cm
\noindent
(iii) We know that black holes emit 
a thermal spectrum and have temperature
and entropy by virtue of semiclassical quantum effects
\cite{Hawk74,Hawk75,DeWi09}. A full theory of
quantum gravity should account for this and should tell us whether or not
the black hole evaporation process comes to an end \cite{Vilk06}.
\vskip 0.3cm
\noindent
(iv) The manifestly covariant theory leads to the detailed calculation
of physical quantities such as cross-sections for gravitational
scattering of identical scalar particles, scattering of gravitons by
scalar particles, scattering of one graviton by another and
gravitational bremsstrahlung \cite{DeWi67c}, 
but no laboratory experiment is in sight for these effects.
\vskip 0.3cm
\noindent
(v) {\it The ultimate laboratory for modern high energy physics is the whole
universe}. We have reasons to believe that either 
we need string and brane theory with
all their (extra) ingredients, or we have to resort to 
radically different approaches such as, for example, loop space
or twistors, the latter two living however in isolation with respect to
deep ideas such as supersymmetry and supergravity (but we acknowledge
that twistor string theory \cite{Witt04} 
is making encouraging progress \cite{Maso05,Toma08}). 

Although string theory may provide a finite theory of quantum gravity
that unifies all fundamental interactions at once, its impact on particle
physics phenomenology and laboratory experiments remains elusive. Some key
issues are therefore in sight:
\begin{enumerate}
\itemsep=0pt

\item  What is the impact (if any) of Planck-scale physics on cosmological
observations \cite{COBEWMAP}?
\item  Will general relativity retain its role of fundamental theory, or
shall we have to accept that it is only the low-energy limit of string or
M-theory?
\item  A\-re re\-nor\-ma\-li\-za\-ti\-on-group me\-thods a 
vi\-ab\-le way to do non-per\-tur\-ba\-ti\-ve
quantum gravity \cite{Reut98,BoRe02}, after the recent discovery of a
non-Gaussian ultraviolet fixed point \cite{LaRe02,ReSa02,LaRe05} of the
renormalization-group flow?
\item  Is there truly a singularity avoidance in quantum cosmology
\cite{Espo05a, Espo05b} or string theory \cite{HoSt90, HoMa95, HoMy95}?
\end{enumerate}

\subsection{Experimental side}

As is well stressed, for example, in Ref. \cite{GrBo10}, gravity is so
weak that it can only produce measurable effects in the presence of
big masses, and this makes it virtually impossible to detect radiative
corrections to it. Nevertheless, at least four items can be brought
to the attention of the reader within the experimental framework.
\vskip 0.3cm
\noindent
(i) Colella et al. \cite{Cole75} have used a neutron interferometer to
observe the quantum-mechanical phase shift of neutrons caused by their
interaction with the Earth's gravitational field.
\vskip 0.3cm
\noindent
(ii) Page and Geilker \cite{PaGe81} have considered an experiment that
gave results inconsistent with the simplest alternative to quantum
gravity, i.e. the semiclassical Einstein equation. This evidence 
supports, but does not prove, the hypothesis that a consistent theory
of gravity coupled to quantized matter should also have the gravitational
field quantized \cite{DeWi62}.
\vskip 0.3cm
\noindent
(iii) Balbinot et al. have shown \cite{Balb10} that, in a black
hole-like configuration realized in a Bose--Einstein condensate,
a particle creation of the Hawking type does indeed take place and
can be unambiguously identified via a characteristic pattern in
the density-density correlations. This has opened the concrete
possibility of the experimental verification of this effect.
\vskip 0.3cm
\noindent
(iv) Mercati et al. \cite{Merc10}, for the study of the Planck-scale
modifications \cite{Amel08} of the energy-momentum dispersion relations,
have considered the possible role of experiments involving nonrelativistic
particles and particularly atoms. They have extended a recent result,
establishing that measurements of atom-recoil frequency can provide 
insight that is valuable for some theoretical models. 

We are already facing unprecedented challenges, where the achievements of
spacetime physics and quantum field theory are called into question.
The years to come will hopefully tell us whether the many new mathematical
concepts considered in theoretical physics lead really to a better
understanding of the physical universe and its underlying structures.

{\bf Acknowledgments} The author is grateful to the 
Di\-par\-ti\-men\-to di Sci\-en\-ze Fi\-si\-che 
of Fe\-de\-ri\-co II U\-ni\-ver\-si\-ty, 
Naples, for hospitality and support. 
He dedicates to Maria Gabriella his work.

\bibliographystyle{abbrv}

\end{document}